\documentclass{article}

\usepackage[utf8]{inputenc} 
\usepackage[T1]{fontenc}    

\usepackage{hyperref}       
\usepackage{url}            
\usepackage{amsfonts}       
\usepackage{amssymb}        
\usepackage{amsmath}        
\usepackage{graphicx}       
\usepackage[numbers]{natbib} 
\usepackage{doi}            
\usepackage{booktabs}       
\usepackage{microtype}      
\usepackage{lipsum}         
\usepackage{algorithmic}    
\usepackage{textcomp}       
\usepackage{xcolor}         
\usepackage{float}          
\usepackage{tikz}           
\usepackage{array}          
\usepackage{tabularx}       
\usepackage{caption}        
\usepackage{subcaption}     
\usepackage[margin=1in]{geometry} 
\usepackage{nicefrac}       

\title{DE-LIoT: The Data-Energy Networking Paradigm for Sustainable Light-Based Internet of Things}

\author{%
\hspace{1mm}Amila Perera \\
Centre for Wireless Communications, \\
University of Oulu, Oulu, Finland \\
\texttt{malalgodage.perera@oulu.fi} \\
\and
\hspace{1mm}Roshan Godaliyadda \\
Department of Electrical and Electronic Engineering, \\
University of Peradeniya, Kandy, Sri Lanka \\
\texttt{roshan.godaliyadda@peradeniya.edu.lk} \\
\and
\hspace{1mm}Marcos Katz \\
Centre for Wireless Communications, \\
University of Oulu, Oulu, Finland \\
\texttt{marcos.katz@oulu.fi}
}

\hypersetup{
pdftitle={A template for the arxiv style},
pdfsubject={q-bio.NC, q-bio.QM},
pdfauthor={David S.~Hippocampus, Elias D.~Striatum},
pdfkeywords={First keyword, Second keyword, More},
}

\begin{document}
\maketitle

\begin{abstract}
The growing demand for Internet of Things (IoT) networks has sparked interest in sustainable, zero-energy designs through Energy Harvesting (EH) to extend the lifespans of IoT sensors. Visible Light Communication (VLC) is particularly promising, integrating signal transmission with optical power harvesting to enable both data exchange and energy transfer in indoor network nodes. VLC indoor channels, however, can be unstable due to their line-of-sight nature and indoor movements. In conventional EH-based IoT networks, maximum Energy Storage (ES) capacity might halt further harvesting or waste excess energy, leading to resource inefficiency. Addressing these issues, this paper proposes a novel VLC-based WPANs concept that enhances both data and energy harvesting efficiency. The architecture employs densely distributed nodes and a central controller for simultaneous data and energy network operation, ensuring efficient energy exchange and resource optimisation. This approach, with centralised control and energy-state-aware nodes, aims for long-term energy autonomy. The feasibility of the Data-Energy Networking-enabled Light-based Internet of Things (DE-LIoT) concept is validated through real hardware implementation, demonstrating its sustainability and practical applicability. Results show significant improvements in the lifetime of resource-limited nodes, confirming the effectiveness of this new data and energy networking model in enhancing sustainability and resource optimisation in VLC-based WPANs.
\end{abstract}


\section{Introduction}
In recent years, wireless smart Internet of Things (IoT) sensor technologies have been rapidly gaining acceptance due to the rapid development of wireless communication networks and the applicability of artificial intelligence and smart algorithms that support their operation. The versatility of wireless sensor networks across various sectors, including automation, health sector, industrial monitoring, logistics, Wireless Personal Area Networks (WPAN), and Wireless Body Area Networks (WBANs) \cite{WPAN, REJEB2022100565}. Notably, mesh networks are gaining traction within low-power IoT sensor WPAN networks. This is due to their distinct attributes, which include reduced energy consumption, extended range facilitated by hop-to-hop communication, improved reliability enabling multi-node communication, and enhanced scalability permitting seamless addition of new devices without extensive configuration or infrastructure modifications \cite{5350373}. The development of future 6G systems is driven not only by  performance goals but also sustainability aspects have to be considered. The need to integrate green and energy-efficient solutions, including concepts such as green IoT and zero-energy IoT, becomes increasingly important \cite{BENHAMAID2022103257,9319211}. In addition to conventional challenges faced by wireless sensor nodes, such as ensuring robust connectivity, energy- and spectral-efficient operation, low-cost production, and ongoing maintenance costs due to limited battery life in designs, there is a need to consider environmental challenges. These challenges include electronic waste generated at the end of the IoT life cycle, especially given the massive number of connected devices.

In order to extend the operational lifespan of IoT devices by improving battery efficiency, the integration of technologies such as Energy Harvesting (EH) proves essential \cite{ACM_EH}. Moreover, EH technologies have gained attention for their potential to provide a sustainable and dependable power source, especially for low-power devices in remote or challenging-to-access locations \cite{MAZUNGA2021e00720}. For indoor applications, Photovoltaic (PV) based EH outperforms competing technologies in resource accessibility and yield \cite{EHcomparison}. In IoT networks that feature EH-based nodes, situations may arise where harvesters accumulate more energy than the storage unit can accommodate. This scenario is particularly common in compact nodes characterised by limited storage capacity. Consequently, the harvesting system is commonly deactivated due to inadequate Energy Storage (ES) availability. From a network perspective, this situation represents sub-optimal utilisation of energy harvesting resources. Additionally, the energy harvesting performance of technologies such as PV based energy harvesting can be reduced due to the lack of direct illumination, such as shadows or blockages, in real-world scenarios.
Technologies such as Wireless Power Transfer (WPT) can be utilised to enhance the overall energy harvesting efficiencies of the network. Notable WPT technologies, including Simultaneous Wireless Information and Power Transfer (SWIPT) and Optical Wireless Power Transfer (OWPT), can be given as examples \cite{HAN2017555,ALAMU2023101983,ASHRAF2022100630}.
Among these, OWPT has gained considerable attention recently as a result of its remarkable features, which enable long-distance and highly efficient power transfer \cite{Laser}. The optical signals also exhibits immunity to Electromagnetic Interference (EMI)\cite{Interferences}. 

In Radio Frequency (RF) IoT WPAN technologies such as Bluetooth Low Energy (BLE) and Zigbee, when used in dense WPAN mesh networks, challenges such as collisions, interferences, and bandwidth contention can arise. As a result, the power consumption of the nodes may increase \cite{park2016adaptive}.
IoT based on Optical Wireless Communication (OWC) is emerging as a promising solution to address the challenge of accommodating new IoT applications within the already congested RF spectrum \cite{OpticalIoT}. The concept Light based Internet of Things (LIoT) has demonstrated indoor visible light's unique potential to serve a dual purpose. This involves enabling Visible Light Communication (VLC) while concurrently providing the means to energise the energy-autonomous nodes through PV energy harvesting. In consequence, this concept provides the possibility of self-sustaining nodes with the intrinsic sustainability of an infinite lifespan. LIoT offers several benefits, including the utilisation of uncrowded optical spectrum for communication, resilience against EMI, virtually limitless spectrum reuse potential, inherent physical security, battery-free and maintenance-free operation, as well as the opportunity to repurpose existing indoor lighting infrastructure \cite{9417484}. However, it is essential to consider that VLC channels exhibit volatility due to factors such as shadowing and multipath scenarios. These conditions can cause fluctuations in signal strength and quality, making it challenging to maintain stable and reliable communication links in indoor VLC systems \cite{Haas:20,VLCvotality}. 
LIoT technology can be employed for applications that necessitate continuous indoor ambient illumination, such as supermarket product labels and storage labels \cite{s21238024}. 
Given that many components such as harvesters, transmitters, displays, and sensors needed for LIoT designs can be fulfilled using sustainable electronics technologies such as Printed Electronics (PE), the prospect of future IoT designs becoming more environmentally friendly is promising \cite{9536579,s21238024}. PE components offer advantages such as rapid manufacturing, decreased material requirements, and the utilisation of thin, flexible, stretchable, and biodegradable substrates \cite{jmmp5030089}. This fosters the creation of environmentally friendly and sustainable components and devices with lower manufacturing costs, contributing to the enhanced sustainability and affordable integration of LIoT design. As PE technology evolves, the eventual complete integration of LIoT nodes using PE becomes plausible. Batteryless energy harvesting-based operations and biodegradable components, included in LIoT designs, will operate a sustainable network with lower maintenance costs, contributing to the reduction of e-waste at the end of the operation.

Advancing previously established frameworks, this work proposes a novel architecture: the Data-Energy Networking-enabled Light-based Internet of Things (DE-LIoT). This LIoT-centric network aims to streamline both data and energy distribution within densely configured WPAN systems. Taking advantage of the unique characteristics of LIoT nodes, the network is designed to be self-sufficient, boasting an unlimited operational lifespan. In this proposed network, data networking is achieved through OWC, while energy networking is facilitated by OWPT. 
This approach aims to mitigate the impact of performance on LIoT-based sensor nodes resulting from the volatile nature of VLC channels in dynamic indoor environments, thereby improving both data communication performance and energy harvesting efficiency. 

The contributions of this research work can be summarised as follows.

\begin{itemize}
    \item Introducing DE-LIoT, a novel data and energy network developed for densely populated indoor EH-based IoT sensor applications, aiming to achieve a prolonged operational lifetime. It leverages the directivity characteristics of OWC and OWPT.

     \item Presenting a distributed sensors and centralised controller architecture applicable to indoor illumination scenarios through DE-LIoT. This architecture identifies resourceful and deficit nodes, demonstrating how it enhances communication and energy harvesting capabilities of deficit nodes.
    
    \item Assessing the feasibility of developing a DE-LIoT network through a comprehensive review of the literature and an examination of currently available, suitable technologies.
    
    \item Demonstrating the feasibility of the DE-LIoT concept by designing, implementing, and evaluating a proof-of-concept network. This network utilises energy-autonomous prototype nodes and access points compatible with the DE-LIoT framework.
    
    \item Evaluating the effectiveness of the DE-LIoT concept through a proof-of-concept prototype network. Highlighting the impact of the DE-LIoT network concept on its performance. 
\end{itemize}

In addition, this research promotes environmental sustainability through the use of an alternative reusable optical spectrum for communication and the re-purposing of illumination infrastructure for both data and energy networking. Furthermore, the prototype design emphasises the consideration of commercially available PE components for implementation, highlighting the potential for further enhancement of sustainability in the future.

The rest of the paper is organised as follows. Section 2 compares the previously implemented OWC networks and the differences with the proposed one. Section 3 discusses the use cases, motivation, and comparison with existing technologies. Section 4 outlines the model of the proposed system and introduces the core concepts of data and energy networking. Section 5 presents the challenges involved in designing a data-energy-networking-enabled LIoT. In Section 6, the implementation details of the prototype are discussed. The prototype's performance evaluation is carried out in Section 7. Finally, Sections 8 and 9 involve discussing the analysis of the achieved results and presenting concluding remarks.

\section{Review of previous work on optical wireless sensor networks}

In the context of OWC and networking, researchers have made significant contributions in both theory and practical applications. A survey conducted in \cite{LIGHTnet} introduces the LIGHTNET concept, which unites free space OWC with networking principles. In a different study \cite{openvlc}, OpenVLC platform is presented as a re-configurable OWC networking platform suitable for various applications. Furthermore, \cite{6402861} proposes a VLC platform that utilises Light Emitting Diodes (LED) for both transmitting and receiving purposes, emphasising the potential to further reduce hardware requirements in OWC networks. Table 1 provides a comparative overview of various OWC-based networking approaches. 
\begin{table}[H]
    \centering
    \small 
    \caption{Comparison of implemented optical wireless sensor networks}
    \begin{tabularx}{\textwidth}{|X|X|X|X|} \hline 
         \textbf{Name} &  \textbf{Networking} &  \textbf{Node Powering Method} & \textbf{Applications} \\ \hline 
         Shine \cite{7366937} &  Distributed Multi-hop &  Battery/wired & Indoor and outdoor sensing/IoT \\ \hline 
         EnLighting  \cite{7732989}&  Distributed multi-hop &  Wired & Indoor and outdoor sensing/IoT \\ \hline 
         AQUAE-NET \cite{9662216} &  Centralised single-hop &  Battery & Underwater sensing/IoT \\ \hline 
         LIoT \cite{9417484,s21238024} &  Centralised single-hop &  Harvested energy (energy-autonomous) & Indoor sensing/IoT \\ \hline 
         DE-LIoT (this work) &  Distributed multi-hop with centralised control in both data-energy networking &  Harvested energy (energy-autonomous) & Indoor sensing/IoT \\ \hline
    \end{tabularx}
    
    \label{Previouswork}
\end{table}

In contrast to existing approaches, our proposed network strategically addresses crucial gaps in the existing literature. While current approaches in optical wireless sensor networks face challenges, such as heavy reliance on batteries or wired power connections, necessitating frequent maintenance, incurring higher costs, and consequently limiting their applicability. Our work aims to bridge this research gap by introducing a comprehensive strategy centred on the exclusive utilisation of freely available indoor-harvested energy in nodes. While existing literature predominantly employs integrated optical transmitters solely for data networking purposes, our approach uniquely addresses a research gap by utilising optical transmitters for both novel energy networking and data networking through OWPT.  Furthermore, deviating from the typical approach of acquiring and sending information for post-processing in conventional IoT sensor systems, the proposed network will utilise such information for its autonomous network optimisation, thereby facilitating more adoptive optimised data and energy networking. 

\section{Use cases and evaluation of existing technologies}

The DE-LIoT technology's application primarily aims to provide sustainable, energy-autonomous intelligent smart sensing across distributed LIoT integrated nodes within premises. Applications requiring smart sensing in illuminated environments, such as laboratories, museums, storage areas, supermarkets, healthcare facilities, and similar settings, are suitable for deployment. In order to achieve these goals, engineering requirements need to be fulfilled, such as low implementation and operation costs, reliable communication, ultra-low power consumption to extend the lifetime, and importantly, EH optimised to sufficiently power smart sensing and processing while ensuring autonomous operation. While current commercialised energy autonomous battery-free technologies for the same purpose, such as passive Radio-Frequency Identification (RFID)  approaches, partially meet these requirements, they exhibit sub-optimal performance in certain areas due to inherent characteristic flaws of the RF spectrum. Despite robust non-line-of-sight (NLOS) communication, the sub optimal EH performance of RF significantly challenges the operation of RF-based nodes as energy-autonomous smart sensing devices. The requirement for dedicated RF transmitters for data and energy transmission undermines sustainability \cite{powercast}. Additionally, RF nodes reliability can be compromised by signal interference from metals and liquids \cite{RFID}. RF based communication also raises security and EMI issues for certain applications, requiring complex countermeasures that increase system complexity and limit its use \cite{Trends_in_Supply_Chain_Design_and_Management_2009, EMI}. 

On the other hand, the use of DE-LIoT, which operates in the visible light spectrum in the aforementioned applications, offers unique advantages of VLC, including immunity to EMI, as previously discussed. Moreover, one of the main highlighted advantages of the LIoT technology is the high efficiency of indoor PV-based EH, which enables a greater energy budget for the energy-autonomous nodes. The increased availability of the energy budget under indoor illuminated conditions allows for the powering of extensive real-time monitoring of node conditions (e.g., sensors detecting changes in temperature, humidity, etc.), intelligent data analysis (e.g., predicting expiration dates and degradation, etc.), the ability to power interactive displays and indicators (e.g., enabling information to be dynamically updated to fit application requirements), and enhanced communication across nodes. Collectively, these improvements enhance the quality of service for LIoT applications. The LIoT applications can also benefit from high-accuracy VLC indoor positioning, which is an essential feature for the scope of use cases \cite{localization, VLC_posi}. The integration of Optical Access Points (OAPs) within existing lighting frameworks facilitates the creation of a distributed sensor network and centralised controller architecture within indoor premises. This supports network communications, node energisation, and premises illumination using a singular infrastructure. Consequently, this integration enhances the sustainability of the application by optimising the use of existing resources.

\section{Model of the proposed data - energy networking system}

\begin{figure}
    \centering
    \includegraphics[width=0.45\linewidth]{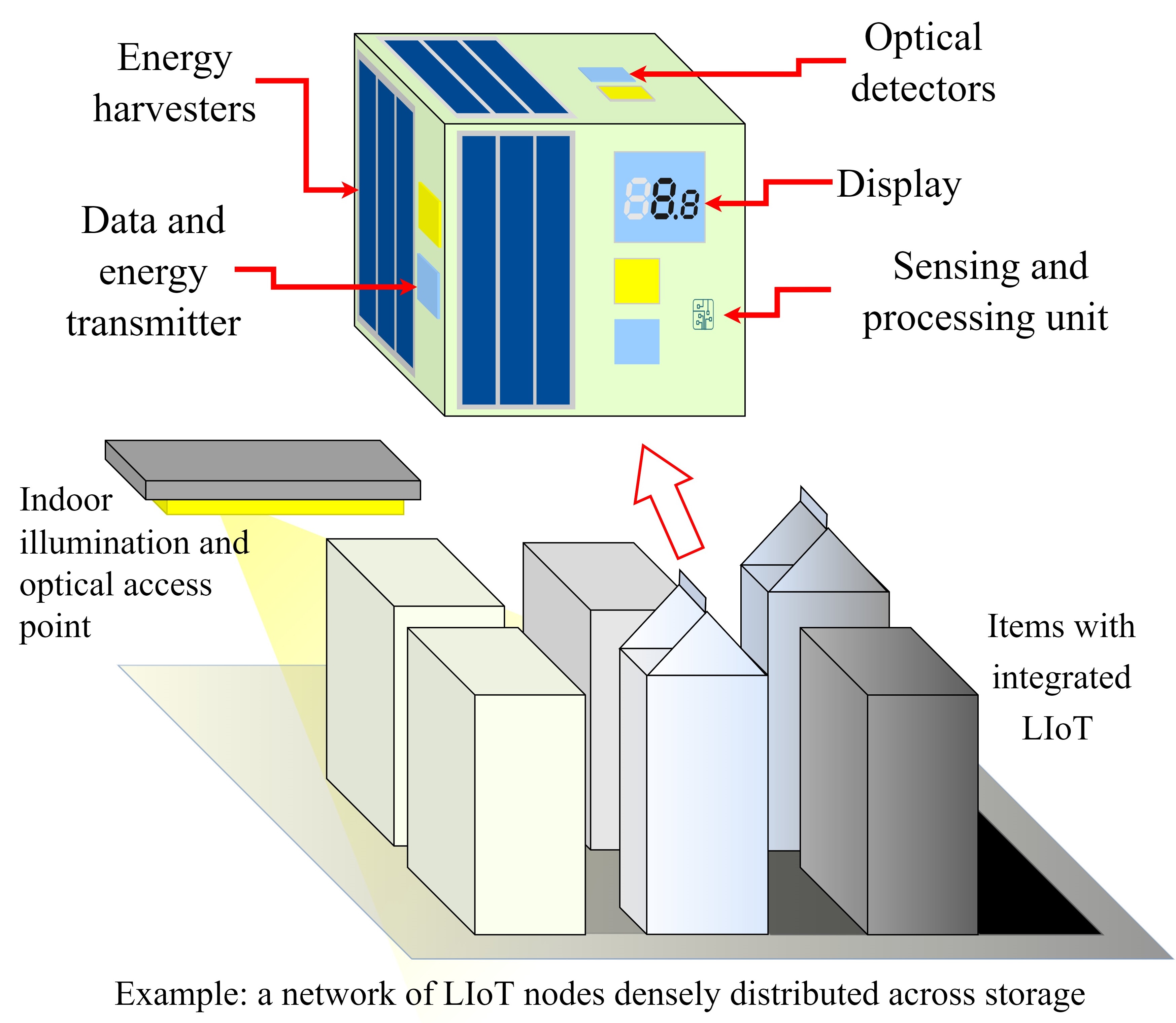}
    \caption{Essential components and typical use case scenario for DE-LIoT.}
    \label{components_application}
\end{figure}

The DE-LIoT node typically encompasses energy harvesters for energy harvesting, a transmitter for data and energy transmission, an optical detector for optical signal detection, and a sensing and processing unit for handling sensing and processing tasks. The essential components for DE-LIoT, along with the typical nature of applications, are illustrated in Figure \ref{components_application}. This proposed data-energy network demonstrates enhanced functionality in premises with continuous illumination, substantial LIoT node density, and dynamic indoor activity.   

In the proposed WPAN architecture, an OAP functions as both a source of indoor illumination and a central controller for all nodes located within its coverage area. The nodes can be categorised as primary or secondary nodes based on the intensity of illumination they receive from the OAP. Primary Sensor Nodes (PSN) located closer to the OAP in areas with better reception conditions in the indoor channel benefit from higher levels of illumination, which ensures efficient energy harvesting and higher signal levels for communication. Conversely, Secondary Sensor Nodes (SSN), either located farther away from the OAP or experiencing unfavourable optical channel and propagation conditions, such as NLOS scenarios, receive diminished illumination. Consequently, these SSNs encounter challenges in both maintaining quality optical signal reception and effectively harvesting energy. Due to the time-varying nature of optical channel volatility, the assignment of PSN and SSN roles can change over time depending on the rate of channel volatility.

In order to address this scenario, a strategic approach is employed in which PSNs, in addition to their primary communication role, are configured to intermittently activate their optical transmitters and transmit their excessively generated energy during a given interval. This will improve the EH performance of these SSN. This concept can accordingly be termed "Energy Spilling", as PSNs gather surplus energy beyond their immediate needs and storage capabilities, which they can then share with neighbouring nodes. 

\subsection{Energy-Data relay nodes}

\begin{figure}[]
    \centering
    \includegraphics[width=0.64\linewidth]{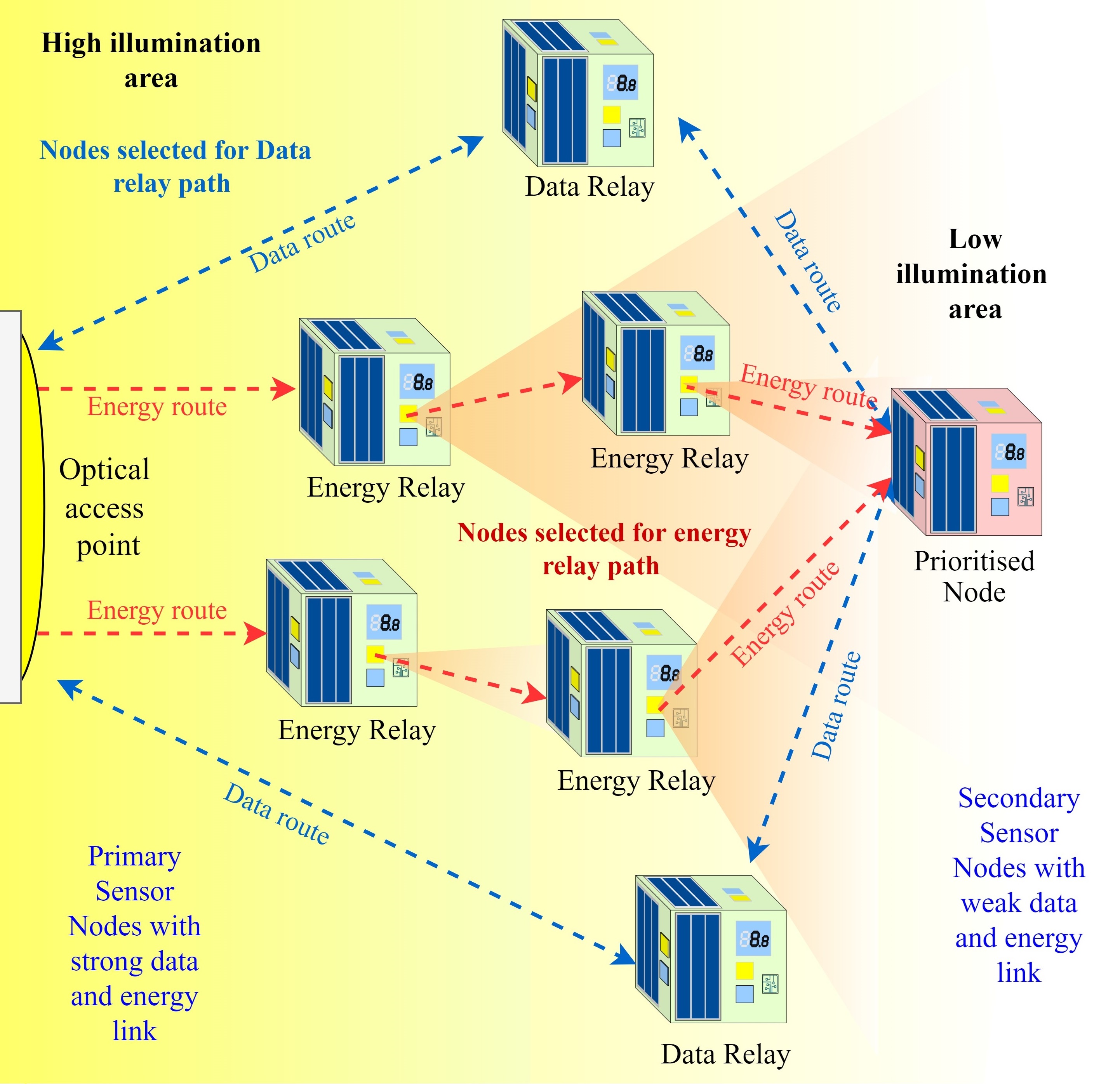}
    \caption{After the OAP prioritises a SSN under low illumination or poor VLC channel conditions, the remaining PSN can be assigned by the OAP to function as energy and data relay nodes. This is achieved by directing data and energy routes towards the prioritised node. This approach aims to enhance the transmitted optical power directed towards the prioritised node.}
    \label{Priritized}\end{figure}

Similar to relay operation in conventional IoT networks, the concept of data relay mode operation can be employed to overcome the challenge of low signal strength between SSNs and the OAP. Through this approach, a PSN assumes the role of an intermediary data routing node, effectively functioning as a "Data Relay". This enables communication between the OAP and an SSN. In environments where nodes are distant from the OAP or encounter NLOS conditions due to dynamic environmental factors such as shadowing or blockage, signal fluctuations may occur. In order to mitigate the impact and maintain communication links adaptively, a multi-hop data transfer strategy utilising node-to-node communication can be employed. Furthermore, techniques such as data aggregation can be integrated with this data relay approach to further enhance the efficiency of energy consumption \cite{randhawa_data_2017}. Similarly, an intermediate node engaged in energy transfer can be termed an "Energy Relay". By utilising the directional characteristics of optical wave propagation, these energy relays can operate using a multi-hop approach similar to data relays. Consequently, a single PSN can serve both as an energy and data relay, contributing to the overall network functionality. Additionally, when the OAP detects an energy deficiency or poor VLC channel conditions in an SSN, it can use calculations and predictions to optimise the value of the "Energy Sharing Parameter" for the selected PSN neighbours to the SSN and route them accordingly. Higher values of the Energy Sharing Parameter can be assigned to the PSN with a low duty cycle, prioritising more ETX functionality, while lower values can increase the duty cycle, resulting in increased data networking time. This optimisation aims to create an energy relay and data relay path to improve energy harvesting and data communication for the SSN. This operation will prioritise the SSN, and the selected PSNs for energy relays will transmit their surplus harvested energy. Meanwhile, the nodes chosen for data networking will operate with extended duty cycles, using their surplus energy to facilitate inter-node communication with the prioritised SSN.
In this way, the SSN is expected to have a better communication route through data relays, with increased harvestable optical energy availability due to the increased ETX bursts toward its harvesters.
In order to make the OAP aware of critical DE-LIoT indicators such as energy availability or EH performance, parameters such as the receiving illumination level at various PV cells on the node and the amount of energy collected and stored can be transmitted periodically using advertisement frames to the OAP. Based on this information, the OAP is expected to approximate localisation and assign nearby PSNs to prioritise resources lacking SSNs. The concept of energy and data relays is visually represented in Figure \ref{Priritized}.

\subsection{Networking nodes with data-energy transfer capabilities}
Generally, for a PSN to operate autonomously while sharing excess energy with neighbouring nodes, the following requirements need to be met:

\begin{equation}
E_{\text{Scavenge}} + E_{\text{Store}} > E_{\text{Oper}} + E_{\text{Sense}} + E_{\text{Process}} + E_{\text{Transmit}} \label{eq}
\end{equation}

where $E_{\text{Scavenge}}$ represents the energy scavenged from the environment, $E_{\text{Store}}$ is the available energy at the storage unit, $E_{\text{Oper}}$ denotes the normal operational energy requirement of the node, $E_{\text{Sense}}$ signifies the energy consumption due to sensing, $E_{\text{Process}}$ indicates the energy used for data processing of acquired measurements, and $E_{\text{Transmit}}$ represents the amount of energy transmitted from the node. 

In a DE-LIoT  sensor network operating with fixed interval sensing, the sample period $T_{\text{Int}}$ comprises both data network and energy network time periods. In the data network mode, nodes are expected to operate with both sensors and transceivers turned on. This allows the node to sense and transmit data acquired from the sensors, while also being able to listen to incoming signals. The data networking mode is pertinent to the data relay operation in the DE-LIoT network.

\begin{figure}[]
\centering
\includegraphics[width=0.88\linewidth]{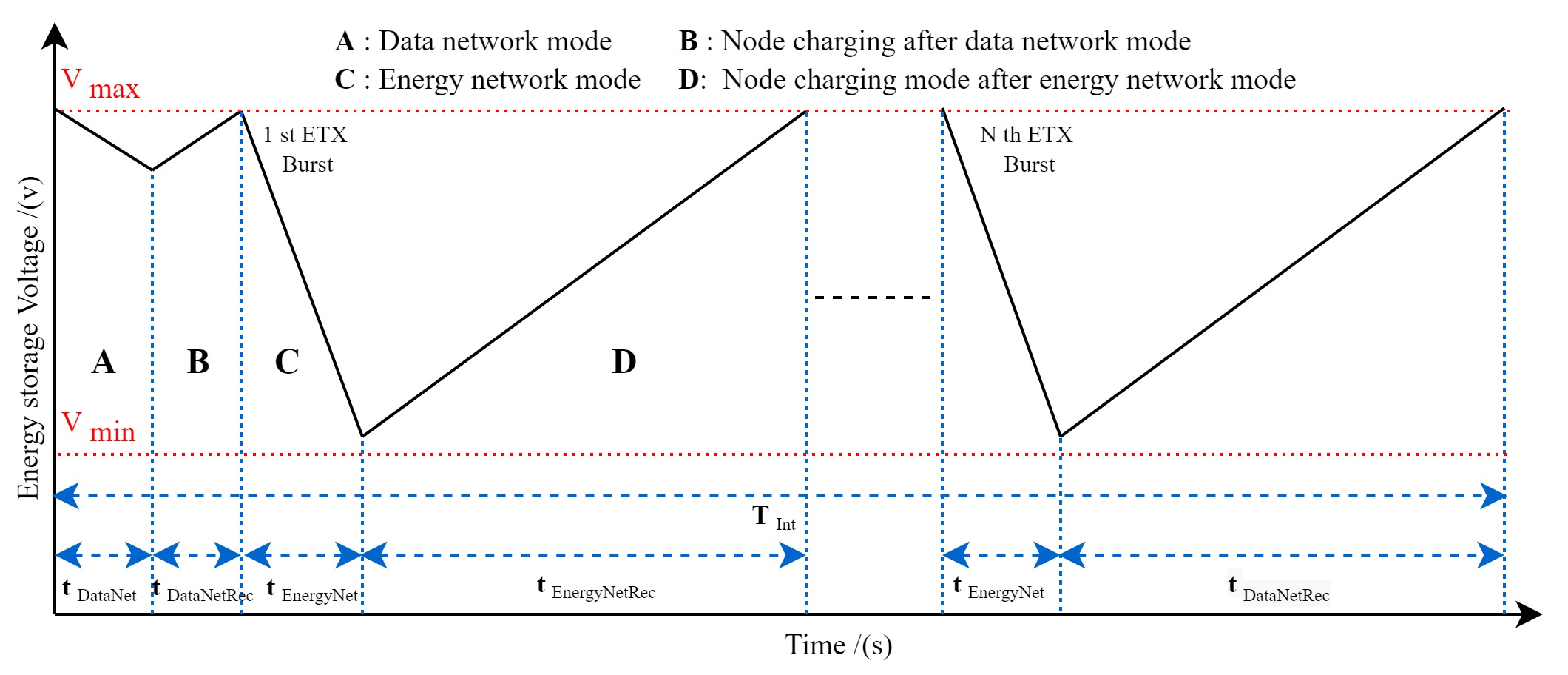}

\caption{Variations in internal ES voltage occur with different time parameters for PSN. 
}
\label{internal_Voltage_var_PSN}
\end{figure}

During the energy networking mode, all nodes are anticipated to power off their sensors and transceivers, prioritising the use of acquired energy for sharing with neighbouring nodes. This energy networking mode is relevant to the energy relay operation in the DE-LIoT network. In order to optimise the use of EH resources within the nodes, it is advisable to establish an ES level range that maximises energy sharing. Voltage levels of the ES can be employed for this purpose, especially when dealing with high-power-density ES devices. Accordingly, specific values for maximum voltage ({\textit{V}}\textsubscript{\textit{max}}) and minimum voltage ({\textit{V}}\textsubscript{\textit{min}}) can be determined to serve this objective. In order to enhance the EH rate, nodes can transition to a low-power-consuming sleep mode, leading to surplus energy generation. 
After a successful data network session, nodes can transition to a low-power sleep state to recover the energy consumed during the session. Once the node reaches the \textit{{V}\textsubscript{\textit{max}}}  threshold and its EH capacity is saturated after the data network mode, the PSN can autonomously initiate energy transmission (ETX). Alternatively, the OAP can also initiate the ETX based on its available information about PSN's  ES level. As the voltage decreases to \textit{{V}\textsubscript{\textit{min}}}, the PSN can switch to a low-power mode, intensifying surplus energy harvesting to expedite a return to \textit{{V}\textsubscript{ \textit{max}} }. This approach enables the PSN to execute a series of \textit{N} ETX bursts, serving as the energy sharing parameter , synchronised with the intervals between sensing periods. The entire intermittent process can be visualised in the diagram shown in Figure \ref{internal_Voltage_var_PSN}. Here, $t_{\text{DataNet}}$, $t_{\text{DataNetRec}}$, $t_{\text{EnergyNet}}$, and $t_{\text{EnergyNetRec}}$ represent, respectively, data networking time, voltage recovery time after data networking, energy networking time, and voltage recovery time after energy networking.

For above discussed system, the time interval can be expressed as follows.
\begin{equation}
T_{\text{Int}} = t_{\text{DataNet}} + t_{\text{DataNetRec}} + N(t_{\text{EnergyNetRec}} + t_{\text{EnergyNet}}) , \label{eq2}
\end{equation}

In a DE-LIoT design focused at minimising energy consumption by deactivating transceiver circuits, the time period \textit{{t}\textsubscript{ \textit{DataNet}}} is the only time a node can interact with communication functions within a given \textit{{T}\textsubscript{\textit{Int}}} period. Based on the expression of  duty cycle for low power consumption of IoT \cite{gotzhein_duty_2020}
\begin{equation}
{r_{DutyCycle}} = {t_{Active} \over t_{DutySchedule} }, \label{eq3}
\end{equation}
Based on (\ref{eq3})
\begin{equation}
{r_{DutyCycle}} = {t_{DataNet} \over T_{Int} }, \label{eq4}
\end{equation}
from (\ref{eq2}) and (\ref{eq4})
\begin{equation}
{r_{DutyCycle}} = 1 - \Biggl({t_{DataNetRec} \over T_{Int} }\Biggr) - N\Biggl({t_{EnergyNetRec} + t_{EnergyNet} \over T_{Int} } \Biggr). \label{eq5}
\end{equation}
Therefore, it is important to select the optimum \textit{$N$} for the PSNs to provide maximum energy transmission while minimising outage time by ensuring sufficient \textit{$t_{\text{DataNet}}$}. This allows the PSN to engage in data networking when an OAP requests it or when a SSN requests it without falling into a low-power sleep state when required. Additionally, by considering the current level of illumination and the characteristics of the VLC channel, the OAP can dynamically adjust the value of \textit{$N$} to optimise the balance between energy networking and data networking accordingly.

The scavenged energy \(E_{\text{Scavenge}}\) during a time interval \(T_{\text{Int}}\) under static channel conditions by an SSN with \(M\) energy harvesters in the presence of \(K\) PSNs can be expressed as

\begin{equation}
E_{\text{Scavenge}} = \sum_{i=1}^{M} \left( \sum_{j=1}^{K} N_{j} P_{\text{Receive}(i,j)} t_{\text{EnergyNet}(j)}\eta_{\text{Conv}(i)} + P_{\text{Illum}(i)} T_{\text{Int}} \eta_{\text{Conv}(i)} \right) \label{eq6},
\end{equation}

where\textit{ {N}\textsubscript{\textit{(j)}}} is the number of ETX bursts from the $j^\text{th}$ PSN, \textit{{P}\textsubscript{Receive\textit{(i,j)}}} is received optical power per ETX burst from  $j^\text{th}$ PSN by $i^\text{th}$ harvester,  {\textit{t}}\textsubscript{EnergyNet(j)} is energy networking time of $j^\text{th}$ PSN,  \textit{{P}\textsubscript{Illum\textit{(i)}}} is received indoor optical power from OAP illumination by $i^\text{th}$ harvester and  {{$\eta$}}\textsubscript{Conv\textit{(i)}} is the optical-to-electrical energy conversion efficiency of the $i^\text{th}$ harvester.

\section{Data – Energy networking enabled LIoT design challenges}

This section of the paper explores into the engineering requirements and the associated design challenges of implementing the DE-LIoT network discussed earlier. It will also evaluate the applicability of commercially available solutions and currently practised approaches in addressing these challenges.

\subsection{DE-LIoT optical access point design requirements}
\begin{figure}[htbp]
\centerline{\includegraphics[width=.68\textwidth]{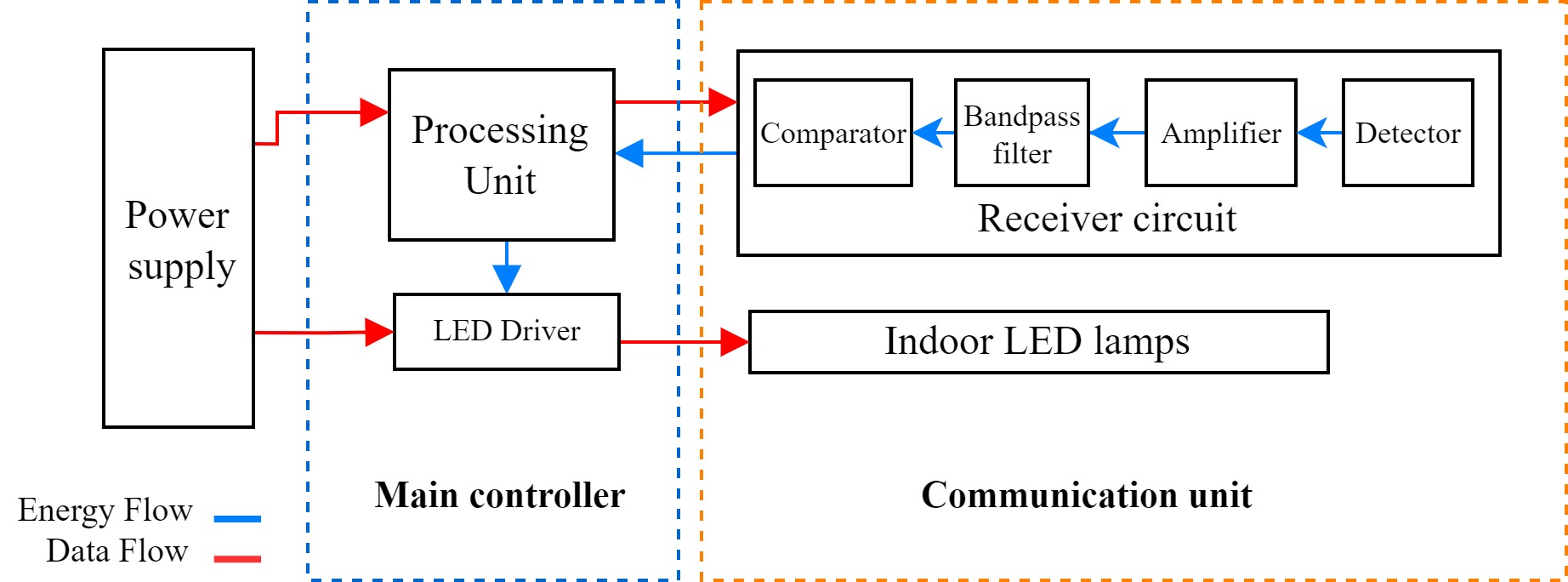}}
\caption{The structure of the DE-LIoT optical access point, capable of performing both indoor illumination and communication purposes.}
\label{OAP_stru}
\end{figure}

In general, an OAP within the system needs to provide adequate illumination to indoor environments while simultaneously ensuring reliable communication performance with the network. For application-specific OAP designs, addressing engineering criteria is essential. This involves adhering to indoor lighting standards, such as Indoor Lighting Standard EN-12464-1, AS/NZS 1680, JIS Z 9125, tailored to each environment and ensuring eye-friendly illumination. The selection between general, local, or localised illumination system types influences the placement of luminaires, depending on the setting \cite{lighting}. The structure of the OAP can be visualised as depicted in the diagram shown  in Figure \ref{OAP_stru}.

\subsubsection{Communication unit} 
The transmitter system of the OAP needs to exhibit efficient illumination performance, along with human friendly flicker-free robust data transmission abilities to facilitate improved detection at the DE-LIoT nodes. IEEE 802.15.7 outlines a variety of features that contribute to establishing eye-friendly VLC networks \cite{6163585}. Furthermore, the illumination distribution of the indoor area should be determined based on the average energy harvesting capabilities of the DE-LIoT nodes. In order to ensure effective illumination, modulation friendly lighting technologies such as high-density LED Chip-On-Board (COB) type light sources can be employed \cite{COBLED}. The necessary power for these sources can be provided using LED driver circuits \cite{Driver}. In order to improve data capture from dispersed DE-LIoT nodes, using multiple optical detectors with angle diversity can enhance communication \cite{IR_multi}.
\subsubsection{Main controller} 
Due to the inherently autonomous nature of this network, the OAP assumes more complex processing tasks to ensure efficient, adaptive network operation based on channel conditions. The OAP is expected to continuously monitor information such as energy availability at the nodes through ES voltages, approximate location details, and channel state using PV voltage levels. Based on this information, it can identify the PSN and SSN according to their associated current channel state. It then routes data and energy relays to SSNs that lack resources by assigning optimised values of \textit{N} to them. In order to support this process, the processing unit should be equipped with the processing capability to manage the required complexity optimisation tasks and concurrently execute parallel operations for efficient operation. Furthermore, machine learning and deep learning techniques can be employed to enhance the OAP's capabilities. Machine learning algorithms can be trained on historical channel state data to predict future channel conditions and also ES levels at the nodes \cite{THOMAS2021102741}. This enables the proactive adjustment of network parameters and resource allocation. Deep learning models can optimise  \textit{N} value assignment based on real-time environmental conditions, thus prolonging the network's lifetime and reducing reliance on external power sources. Given that the OAP can also function as a network gateway, it is crucial to have networking compatibility features in place.


\subsection{DE-LIoT node design requirements}
\begin{figure}[htbp]
\centerline{\includegraphics[width=.52\textwidth]{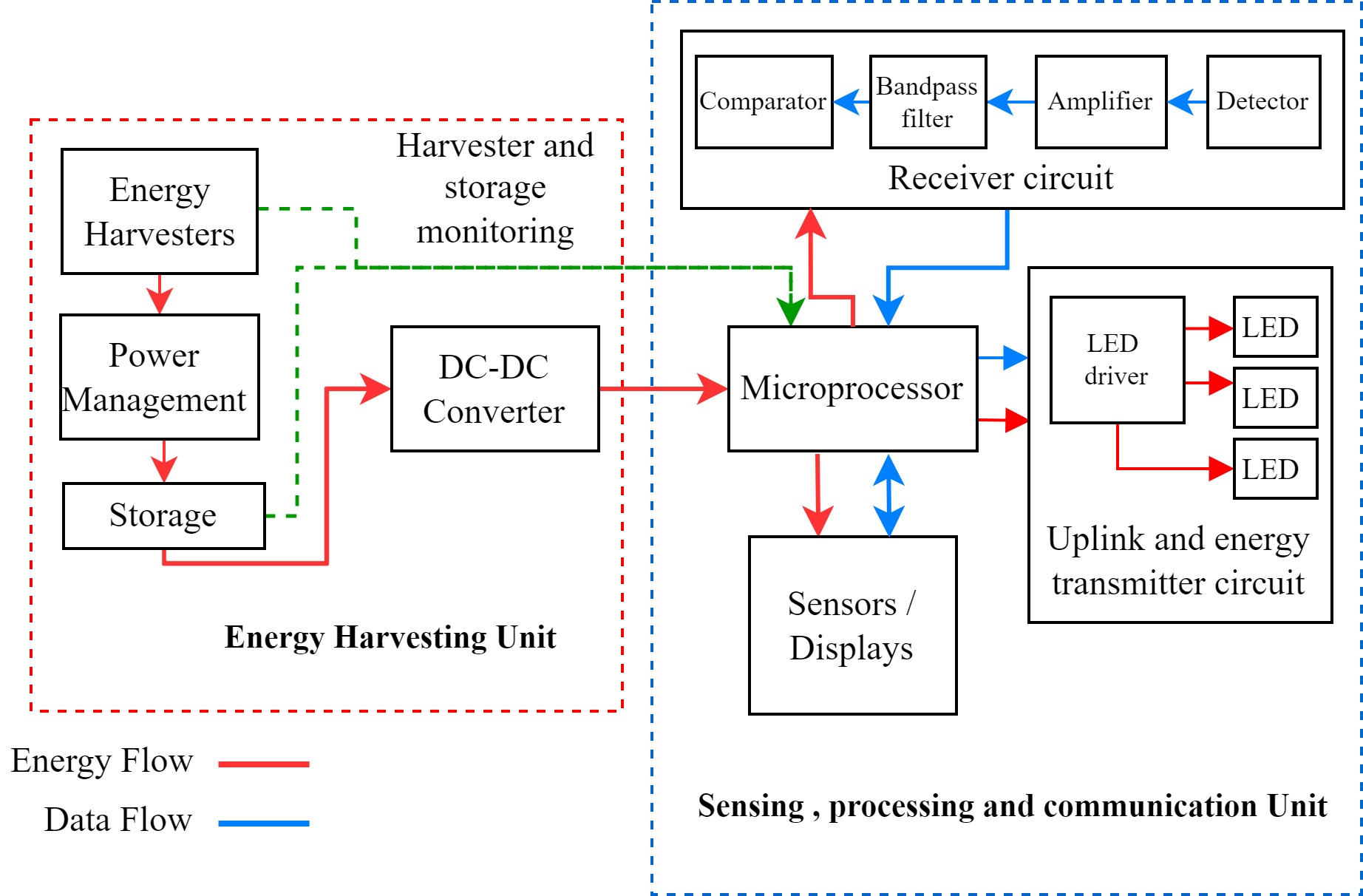}}
\caption{ Structure of DE-LIoT node with energy harvesting unit, along with sensing, processing, and communication units.}
\label{DE_LIOT_STRU}
\end{figure}

The design of DE-LIoT includes an energy-harvesting unit along with sensing, processing, and communication units, as depicted in Figure \ref{DE_LIOT_STRU}. In order to design a DE-LIoT-enabled node, the energy bounds need to follow the condition in (\ref{eq}). In order to fulfil this condition, it is imperative to implement strategies that minimise power consumption during active operations and adopt an EH rate that exceeds the node's average power consumption. Additionally, adept management of intermittent energy availability is essential. These measures are essential to meet the engineering requirements, which depend on the nature of the DE-LIoT node sensing application and the deployment environment. This approach assists in harvesting and storing excess energy for data and energy networking. Achieving this is crucial under standardised illumination conditions defined for application-deployed areas. Additionally, the energy storage unit has to be capable of storing sufficient energy to operate node for extended periods while maintaining the duty cycle required by the application. The design of the optical transceiver on the node, as well as the node's physical geometry, maximum distance between DE-LIoT nodes needs to also support an application-friendly layout once deployed.

\subsubsection{Energy harvesting unit requirements} 
The harvesters of the EHU convert the optical energy of incident optical waves into electrical energy. For this purpose, optical rectennas \cite{yahyaoui_miim-based_2022} or PV cells can be used. In order to ensure compatibility between EH devices and the wavelengths used for data transmission in DE-LIoT networks, it is crucial to select appropriate harvesters. PV cells, for instance, exhibit wavelength responses based on the semiconductor materials employed in their composition \cite{PVPV}. Therefore, for an effective DE-LIoT design, it would be advantageous to employ PV cells equipped with semiconductor materials that respond to the wavelength range designated for both energy and data exchange purposes. Since PV cells typically require substantial surface area of the node, the use of sustainable PE compatible PV technologies, such as Perovskite and Organic Photovoltaics (OPV), can enhance the sustainability of these nodes \cite{perv, opv}. The necessity for a direct LOS illumination in PV-based EH methods could pose a disadvantage when integrated into IoT systems, especially indoors, due to the presence of highly multipath and reflective environments \cite{Haas:20}. A potential solution to address this challenge involves employing a configuration of multiple PV cell arrangements with distinct spatial orientations. Additionally, technologies such as the Shadow-effect Energy Generator (SEG) can be integrated into the DE-LIoT framework. The SEG utilises partially illuminated surfaces of the node to generate energy, expanding the EH capabilities of the system \cite{zhang_energy_2020}. Anti-reflective layer technologies, such as multi-scale array structures, hold the potential to enhance photon absorption in PV cells. This enhancement leads to higher EH efficiencies, particularly when the PV surface is not directly illuminated \cite{li_multiscale_2018}.

In order to enhance the EH efficiency of the DE-LIoT system, load matching operations and adapting EH performance in response to fluctuations can be achieved by using power management sub-circuits. For PV cell-based EHUs, the Power Management Integrated Circuit (PMIC) with Maximum Power Point Tracking (MPPT) can be used \cite{MPPPT}. Generally, these MPPT algorithms have sampling intervals that monitor the voltage variations on the PV terminals. Low-voltage MPPT PMICs often incorporate algorithms such as Fractional Open Circuit Voltage (FOCV) and Hill Climbing (HC). Based on that, the MPPT algorithm will adjust the load matching technique in order to obtain maximum power transfer between PV cell and storage \cite{MPPT}. In these PMIC’s, typically a high performance DC boost converter is used in order to multiply the PV voltage before sending it to the storage. In the context of the proposed DE-LIoT network, the efficient utilisation of rapid optical power surges from the ETX is vital for the SSN under low illumination, as well as the adaptation to indoor illumination fluctuations. Therefore, selecting a PMIC with a higher or adoptive MPPT tracking interval capabilities will be advantageous. 
Furthermore, incorporating features such as a cold start circuit would provide an additional advantage in scenarios where the storage level is completely depleted. The cold start circuit efficiently recharges the storage until it reaches a point where the boost converter and other sub-circuits within the PMIC can operate normally. Compact design features, such as an inductorless design and regulated output (i:e., Low-Dropout regulator (LDO), Buck converter) for the electronic load, offer distinct advantages for creating compact DE-LIoT designs, eliminating the need for additional circuitry.  
The inbuilt low storage level warning signal feature can be useful to inform the MCU of the node so it can minimise the impact of power outage before it occurs. Table \ref{PMIC_T} describes the few commercially available PMICs \cite{epeas-aem1094,NEH2000BY,AnalogDevices2019,epeas2023product}.

\begin{table}
\small
\caption{Commercially available DE-LIoT design friendly PMICs}
\label{PMIC_T}
\begin{tabularx}{\linewidth}{|>{\hsize=1.2\hsize\centering\arraybackslash}X|>{\hsize=0.8\hsize\centering\arraybackslash}X|>{\hsize=0.8\hsize\centering\arraybackslash}X|>{\hsize=0.8\hsize\centering\arraybackslash}X|>{\hsize=1.4\hsize\centering\arraybackslash}X|}
\hline
\textbf{Model} &
  \textbf{MPPT Technique} &
  \textbf{Min MPPT Sampling (s)} &
  \textbf{Input Voltage Range (V)} &
  \textbf{DE-LIoT Friendly Features} \\ \hline
Epeas AEM10941 &
  FOCV &
  5 &
  0.005-5 &
  Low storage warning, Cold start circuit (0.38v), LDO output (80mA) \\ \hline
Analog Devices ADP5091 &
  FOCV &
  2 &
  0.08-3.3 &
  Low storage warning, Cold start circuit (0.38v), LDO output (150mA) \\ \hline
Nexperia NEH2000BY &
  HC &
  0.7 &
  5 (max) &
  Inductorless compact design \\ \hline
Epeas AEM13920 &
  FOCV &
  0.134 &
  0.11-5 &
  I2C communication, Dual harvesting sources, Low storage warning, Cold start circuit (0.275v), Buck output (100mA) \\ \hline
\end{tabularx}
\end{table}

Therefore, the choice of a PMIC can be determined by both the specific needs of the application and the compatibility with other components within the energy harvesting system. 

The selection of the ES for DE-LIoT needs to be made on the characteristics of the charge - discharge nature. Since DE-LIoT devices are expected to collect more energy during short, high-optical power receiving energy-transmitting windows, it is important to employ ES devices that can quickly absorb the available power. Furthermore, considering the increased charge and discharge characteristics of DE-LIoT nodes due to ETX operations, the ability to withstand a potentially large number of charging cycles becomes essential for the design. However, the ES unit needs to have the ability to store sufficient energy to accommodate the rapid changes in power requirements that occur within the MCU unit and the electronics. In real-world scenarios, it is vital to consider non-illuminated periods where PV based EH fails. Thus, storing adequate energy to meet application demands and maintain operations in SSN mode is crucial. Therefore, selecting ES with both higher power density and adequate energy density is crucial. In order to comply with these requirements, ES devices such as super-capacitors and lithium ion capacitors can be used \cite{supercap,yue_charge-based_2020}. Printed electronics-based supercapacitors can further enhance sustainability \cite{printedcap}. According \cite{van_leemput_energy_2023} the minimum required capacitance for energy harvesting-based IoT can be written as

\begin{equation} C_{\mathrm {Min}} \approx \frac {2\left ({\frac {E_{\mathrm{ Peak}}}{\eta _{\mathrm{ PMIC,L}}}+P_{\mathrm{ Leak}}T_{\mathrm{ Peak}}}\right)}{V_{\mathrm {Max}}^{2} - V_{\mathrm {Min}}^{2}},\label{eq7}\end{equation}

where, \textit{{E}\textsubscript{Peak}} is the maximum energy requirement, \textit{{{$\eta$}}\textsubscript{PMIC,L} }is efficiency of transferring energy from the super-capacitor to load, \textit{{P}\textsubscript{Leak}} is leakage power of the super-capacitor and \textit{{T}\textsubscript{Peak}} is the maximum energy consumption period possible for the node. 
In a DE-LIoT node, \textit{{E}\textsubscript{Peak} }is generally expected to coincide with the occurrence of the ETX function, as \textit{{T}\textsubscript{Peak}} equivalent to \textit{{t}\textsubscript{EnergyNet}}.

In order to achieve the necessary voltage regulation for powering the electronics load using the harvested energy stored, a DC-DC converter is essential in the design. As the output voltage of the high power density storage changes with energy depletion, DC-DC converters can be employed to deliver a consistent voltage output to the electronics load. These converters adapt to varying input voltage levels from the storage, ensuring stable power delivery. Typically, high efficiency step down buck converters are used for this purpose.  Moreover, opting to power the electronics through a DC-DC converter rather than utilising the output directly from the PMIC brings an additional benefit. It enables the supply of a higher output current, typically reaching around 300 - 400 mA, exceeding the capacity of the built-in output regulator of the PMIC \cite{TPS62740,ST1PS03}. This will facilitate the use of power-demanding sensors and transmitters with the DE-LIoT node. However, a disadvantage to mention is the power loss that occurs at the DC-DC converter. Choosing a DC-to-DC converter with lower quiescent current and higher efficiency within the desired average output range can minimise power losses during the step-down conversion process. Therefore, the choice of using DC-DC converters depends on the specific type of application of the DE-LIoT. 

\subsubsection{Sensing, processing and communication unit requirements}
Given the power constraints inherent in DE-LIoT nodes, selecting the microcontroller unit (MCU) for the DE-LIoT node involves a trade off between the required computational complexity for the sensing operation and the node's power consumption.  Big-O notation can be employed to determine the time complexity of an algorithm which is related to the computational complexity of an application. Time complexity measures how long an algorithm takes to run as a function of the input length. For IoT applications, algorithms that execute in constant time, such as sensor readings, are {\textit{O(1)}}. In contrast, tasks requiring averaging are {\textit{O(n)}}, with execution time increasing linearly with input size. Thus, complexity progresses from {\textit{O(1)}} to {\textit{O(n)}} \cite{7122861, Bigo, bioengineering10060703}.

The majority of general-purpose microcontroller designs extensively use complementary metal oxide semiconductor (CMOS) technology. For a more detailed analysis, the dynamic power dissipation of a processor based on CMOS technology can be expressed as

\begin{equation}
P_{Dynamic }  \alpha  V_{Supply}^{2} f_{Operating} , \label{eq8}
\end{equation}

where , \textit{{V}\textsubscript{Supply}  }is supply voltage and ,\textit{ {f}\textsubscript{Operating} }is operating frequency of the MCU. Hence, it is crucial to employ an optimised clock frequency that aligns with the MCU's functions, while maintaining a sufficiently low supply voltage that ensures smooth operation without any performance degradation \cite{mccool_structured_2012}. Voltage optimisation can be achieved by employing Near-Threshold Computing (NTC),  Sub-Threshold Computing techniques \cite{Sub,7864441}. NTC involves operating transistors at or near their threshold voltage, which effectively decreases dynamic power dissipation. Soft intermittent task execution, suitable for intermittently EH powered IoT devices such as DE-LIoTs, can benefit from an ultra-low-power MCU that features minimal power consumption during sleep mode \cite{intermitt2,intermittent}. This allows for rapid energy recovery following task execution. Additionally, the presence of non-volatile memory becomes crucial for storing critical data during unexpected power interruptions, which are expected in the context of DE-LIoT. The integration and capacity of Electrically Erasable Programmable Read-Only Memory (EEPROM) flash memory within the MCU holds special significance in this context \cite{NVM}. Peripheral components such as Analogue-to-Digital Converters (ADC), General-Purpose Input Output (GPIO), and Real-Time Clock (RTC) play a vital role in facilitating the operation of the DE-LIoT sensor node. Numerous MCU designs and architectures, including ARM and AVR, are available to support these functionalities. 
Detailed comparative information on the supply voltage, clock speed range, sleep power consumption, and non-volatile memory concerning currently available low-power MCU is presented in Table \ref{UL_MCU} \cite{e-peas_edms105n_2023,ATmega328P_Datasheet,STM32L010F4}.

\begin{table}[]
\small 
\centering
\caption{Commercially available low power microcontroller units}
\label{UL_MCU}
\begin{tabularx}{\linewidth}{|*{6}{>{\centering\arraybackslash}X|}}
\hline
\textbf{Micro-controller} & \textbf{Type} & \textbf{Input Voltage (V)} & \textbf{Min. Sleep Current ($\mu$A)} & \textbf{Clock Freq. Range (Internal)} & \textbf{Non-volatile Memory} \\
\hline
Epeas EDMS105N & ARM (32-bit Cortex M0) & 1.8-3.3 & 0.34 & 32kHz-24MHz & 256kB Flash \\
\hline
Atmega328P & AVR (8-bit) & 2.7-5.5 & 1 & 128kHz-8MHz & 32kB Flash, 1kB EEPROM \\
\hline
STM32L010F4 & ARM (32-bit Cortex M0+) & 1.8-3.3 & 0.23 & 32kHz-32MHz & 16kB Flash, 128kB EEPROM \\
\hline
\end{tabularx}
\end{table}

An advantage of DE-LIoT design is that both signal transmission and energy transfer utilise the same hardware components and setup. This means that the same hardware can be used to perform both functions. The data encoding and modulation can be handled by the MCU, with the output being delivered through GPIO pins at Transistor-Transistor Logic (TTL) voltage. LEDs can be employed as transmitters in DE-LIoT designs. In order to address concerns related to state switching delay and the relatively low power delivery efficiency of GPIO pins on low-power MCUs, Bipolar Junction Transistors (BJTs) that are compatible with TTL voltage levels can be used. 
Techniques such as angular diversity transmission can be used to achieve multi-directional transmission \cite{7857700}. In order to control the directivity of the transmission, LEDs with different half-angle intensities can be employed.

 The selection of the type of LED for DE-LIoT nodes depends on the application of the network. LED wavelengths that fall within the visible spectrum for human or animal eyes might cause discomfort when engaged in data and energy networking for indoor applications.  In the recent IEEE 802.11bb standard, the utilisation of the invisible spectrum ranging from 800 to 1000 nm is being considered for communication purposes \cite{9855461}. In the visible wavelength range, the human eye employs scotopic and photopic vision modes, which correspond to varying levels of darkness and brightness in the environment \cite{scotopic}. Thus, for DE-LIoT networks operating in places with human presence, it is recommended to configure the system to use wavelengths outside the range of the human eye sensitivity. 
On the other hand, the conversion of a value from wavelength to energy can be obtained from \cite{mooney_get_2013}.
\begin{equation}
E = h\nu = \frac{hc}{\lambda},\label{eq9}
\end{equation}
where is \textit{h} plank constant, $\nu$ is frequency, \textit{c} is speed of in vacuum and \textit{$\lambda$} is the wavelength of the wave. From (\ref{eq9}) it can be seen that higher wavelength regions, such as the IR region, carry lower energy compared to the visible and ultra violet regions. Therefore, approaches such as switching to the wavelength range based on time or detecting the presence of humans from the OAP can be used to optimise both data energy networking and create eye-friendly living conditions within the premises. 

The data-receiving unit for the DE-LIoT node can be designed by using a PV cell or a typical Photo-diode (PD) coupled with a transimpedance amplifier. Subsequently, a filtering approach can be applied to extract the data signal \cite{s21238024}. Furthermore, by adopting a low-power design approach for the receiver circuit and utilising a correlator \cite{wux}, it can also serve as a wake-up receiver (WuRX) for the DE-LIoT node. In order receive signals from multiple directions, angle diversity receiver techniques can be employed for the design \cite{HAAS2020443}. In \cite{Sphere5}, a proof-of-concept prototype of an omnidirectional optical antenna is presented, featuring a spherical structure with multiple detectors. Fully printable organic PD  also have the potential to be seamlessly integrated onto node surfaces, thereby contributing to sustainability \cite{PrintedPD}.




In indoor environments where optical data or energy is transmitted over short distances utilising a Lambertian light source such as an LED, the optical path loss can be described by the following equation:

\begin{equation}
L = \frac{(m+1)A}{2\pi D^2} \cos^m(\alpha) \cos(\beta),
\end{equation}

where $m$ denotes the Lambertian order, $A$ represents the detector area, $D$ is the distance between the LED and the detector, $\alpha$ is the irradiance angle, and $\beta$ is the incidence angle. These angles indicate the orientation of the LED relative to the line of sight and the orientation of the photodetector towards the LED, respectively \cite{pathloss}. Therefore, maintaining efficient optical data and energy transmission requires consideration of key factors such as the optimal distances between nodes and the use of optical transmitters with a suitable Lambertian order for light emission, tailored to the application's layout. Once developed, printable organic LED technology can be used for compact, sustainable LED integration in nodes \cite{printedOLED}.

In order to keep track of the EH rate and the current stored energy, internal monitoring is crucial \cite{stash}. Having this awareness of information allows the node to predict and schedule its intermittent operations accordingly. This enables the node to match its energy consumption and generation effectively, thereby preserving its energy autonomy. The internal monitoring of the DE-LIoT needs to be carried out in a manner that the sensing operation does not impact energy harvesting or storage. Additionally, the sensing circuitries should avoid consuming extra power when not in use. For this purpose, isolation techniques such as using switch between sensing circuit and the source can be implemented. 

For application-specific measurement purposes, sensors designed with low-power designs and technologies, such as printed electronics or conventional methods, can be employed.  In \cite{printedsensors,YUE2023109983} the authors discuss the applicability of low-power, low-cost printed gas sensors in various applications, including food spoilage, air quality monitoring, health tracking, and hazardous gas detection. These sensors can identify emitted gases such as ammonia for protein-rich products and ethylene to assess the ripeness of fruits and vegetables, as well as the integrity of the packaging by detecting atmospheric gases. Printed gas sensors also enable monitoring of environmental quality by measuring levels of atmospheric and hazardous gases. Additionally, printed gas sensors can be employed in health monitoring by analysing exhaled breath conditions and detecting emitted odours. 
On the other hand, low-power, IoT-friendly conventional sensors, such as pressure, temperature, humidity, and accelerometers, with power consumption in the nano to micro-Watt range, can be seamlessly integrated into DE-LIoT node setups \cite{tang_energy_2018}. In potential DE-LIoT applications across diverse areas such as logistics, healthcare, food, and environmental monitoring, substantial benefits can be realised through the utilisation of low-power sensors. These low-power sensors play a crucial role in minimising power consumption, thus contributing to the maintenance of energy-autonomous network operations.

\subsection{Communication }
Due to the fact that DE-LIoT applications require lower data rates and implementation designs are optimised for low-power characteristics with limited computational power, single-carrier modulation with reduced complexity can be used for communication. Hence, for this purpose, low-cost digital modulations compatible with intensity modulation and direct detection, such as On-Off Keying (OOK), Pulse Width Modulation (PWM), Pulse Position Modulation (PPM), and Pulse Amplitude Modulation (PAM), can be employed \cite{yang_composite_2020}, \cite{tsonev_avoiding_2014}. 
In order to establish communication with multiple DE-LIoT devices in a shared indoor environment, it is necessary to employ Multiple Access (MA) techniques that complement the nature and specific requirements of the application. According to \cite{eltokhey_multiple_2019}, various MA techniques can be applied for VLC network compatibility, including Time Division Multiple Access (TDMA), Space-Division Multiple Access (SDMA) and Optical Code-Division Multiple Access (O-CDMA). Furthermore, in \cite{Energy-based} the authors propose an adaptive MA framework that focusses on optimising the transmission strategy considering the energy states within a network made up of nodes that rely on energy harvesting. Given that, it is essential to carefully select an appropriate MA technique, taking into account the low-power operation and minimal processing complexity of the DE-LIoT nodes. Compatibility with both node-to-OAP and node-to-node communication is essential in making this selection. 

Due to the directional nature of optical communication, interference and collisions are inherently reduced compared to RF communications. However, the centralised configuration of OAPs might lead to inherent congestion and increased opportunities for collisions in the uplinks, which can be resolved. In order to improve the amount of energy available for energy networking while simultaneously reducing energy consumption in data networking and mitigating data collision and interference \cite{huang_evolution_2013}, effective management of access to the shared medium in DE-LIoT systems becomes crucial. This can be achieved by utilising energy-efficient Media Access Control (MAC) protocols. The choice of MAC protocol should be in accordance with the specific nature and requirements of the DE-LIoT application.  
A less densely deployed DE-LIoT network with spaced layout may benefit from Directional-MAC (D-MAC), a MAC protocol compatible with directional antennas and receivers \cite{DMAC}. In general MAC protocols which designed preserve nodes energy can be used for densely deployed DE-LIoT applications. such as sender-initiated Berkeley MAC (B-MAC), receiver-initiated Predictive Wake-up MAC (PW-MAC), asynchronous and locally synchronised Sensor MAC (S-MAC), where all nodes are expected to use a fixed duty cycle. Additionally, Timeout MAC (T-MAC) can be applied in scenarios where protocols involve nodes anticipated to have a dynamic duty cycle \cite{1306496}. 

While OWC systems inherently offer physical security advantages due to LOS communication, it is essential to acknowledge and address potential security risks that may still be present.  In \cite{security} authors identify potential security stages for OWC networks: Before Signal Emission (BSE), During Signal Propagation (DSP), and After Signal Receiving (ASR). In order to mitigate these security threats for DE-LIoT, protocols such as physical layer security (PLS), which is more compatible to energy-constrained systems, and MAC layer security based on IEEE 802.15.7 standards can be employed.  In order to address BSE threats in DE-LIoT networks, strategies such as employing complex modulation schemes and implementing channel hopping can be considered. For mitigating threats DSP, approaches such as optical beam-forming or the utilisation of hybrid RF OWC links can be implemented.

\begin{figure}
\centerline{\includegraphics[width=.5\textwidth]{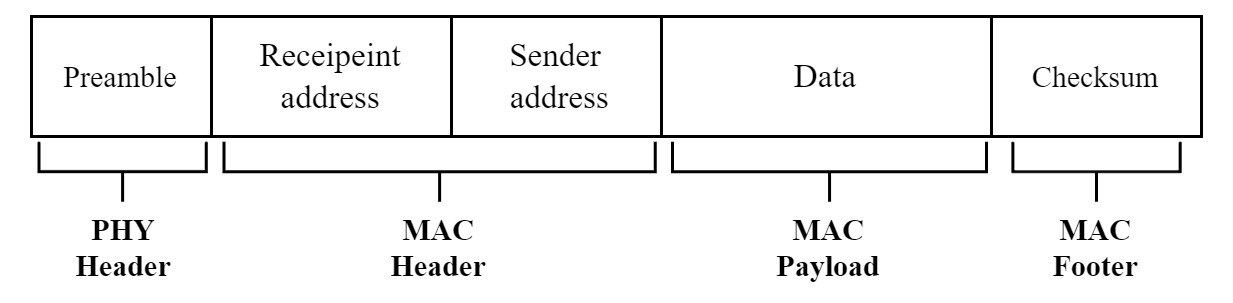}}
\caption{The proposed communication data frame structure for the DE-LIoT network, based on the PHY and MAC layers.}
\label{Prp_data_frame}
\end{figure}

As described in the previous section, sending essential information to enable situational adaptive decisions by the OAP requires periodic transmission of advertisement packets from DE-LIoT nodes. Following the IEEE 802.15.7 standard for short-range optical wireless communication, which defines the physical layer (PHY) and the MAC layer \cite{khan_visible_2017}, the data frame structure for the  node to OAP advertisement, sensor data transmission, and bidirectional OAP communication between nodes can be described as in Figure \ref{Prp_data_frame}.

In here, the preamble can be used for functions such as a wake-up signal, synchronisation, and frame identification purposes. Address bits can be used to contain sender and recipient information, while data bits are used to hold data. For an advertisement frame, these bits could contain information such as storage level and receiving illumination information of the node. In data frames, these can be used to contain sensor data information, depending on the application. For the downlink from the OAP to the node, these bits can be used to convey command information, such as instructions based on the optimal \textit{N} value, duty cycle, and wavelength selection for ETX. The final checksum bits can be used to include checksum methods to detect errors in received data.
\section{Prototype implementation}

This section describes the prototype testbed and framework made to analyse the behaviour of LIoT data - energy networking with the use of commercially available equipment. For the prototype framework, one access point and three DE-LIoT compatible prototype nodes were developed. The selection of components for the prototype was guided by prioritising factors such as commercial availability, cost-effectiveness, integration feasibility with other components, and compatibility with previously developed algorithms. It is noteworthy that this prototype implantation approach may not yield the performance levels attained by the highest efficiency and lowest power-consuming hardware currently on the market. The implementation included support for energy-autonomous operation, idle transmission schemes, and data modulation schemes, all of which were implemented in an approach similar to our previous work \cite{9417484,s21238024}. The prototype system was designed to comply with a potential DE-LIoT application scenario, accommodating suitable indoor-friendly illumination conditions, a sensing application, a sensing cycle, and the physical dimensions of nodes and OAP.

\subsection{Communication and parameter selection for the prototype network implementation }

\subsubsection{Prototype communication protocol for the network}
\begin{figure}[]
    \centering
    \includegraphics[width=1\linewidth]{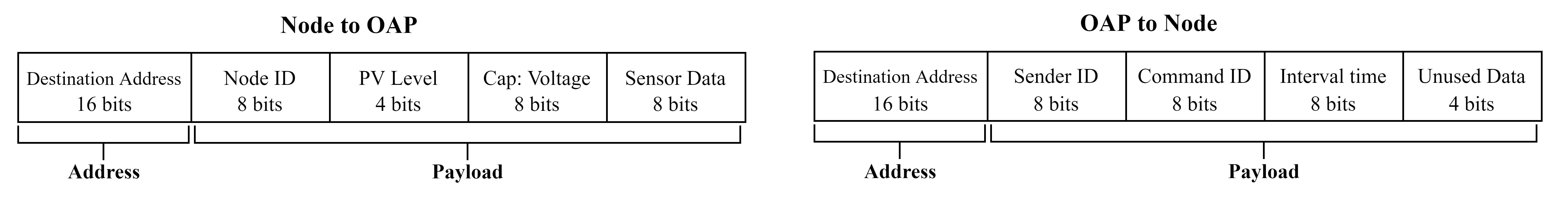}
    \caption{Data frame structure used in the prototype network, where advertisements and data are fused within the same data frame for Node to OAP communication.  }
    \label{Data_frame_PROT}
\end{figure}

The communication protocols employed in the prototype framework are inspired by IR remote control communication protocols commonly found in consumer electronics. Both the OAP and the node prototypes utilise a 44-bit data frame structure for communication. For the implementation of the communication protocol for the prototype, the IR remote version 2.5 library  was used within the Arduino integrated development environment. In the prototype network, the intended communication from the node to the OAP involved concatenating both advertisement and data information into the same data frame. The allocation of bits of data frame is illustrated in Figure \ref{Data_frame_PROT}.

Within this framework, 16 bits are dedicated to the destination address, while the remaining 28 bits carry the payload. In the case of OAP to node communication, the address segment designates the node address for the transmitted payload. Furthermore, the "Sender ID" segment within the data frame signifies the OAP's ID to identify its data. Moreover, a "Command ID" is included to correlate with the necessary command index for the impending task. The data frame also incorporates an "Interval time" segment, housing data pertaining to the latest data collection interval. 
Similarly, the node communication protocol adheres to the same optical communication approach.  When there exists an unobstructed LOS between the OAP and the node, the address pertains to the access point; in contrast, inter node communication employs the nearest node's address. The payload contains the node's own ID in the "Sender ID" segment, allowing the access point to determine the payload's source. It also includes essential data such as PV level, capacitor voltage, and sensor data, all vital for future processing of the suitable \textit{N} at the access point.

The length of the data frame was chosen for the prototype network to support the necessary number of nodes and OAPs while handling sensor, PV, and capacitor readings with adequate precision. Although this choice limits expandability with a large number of nodes under a single OAP and the decimal accuracy of each reading obtained by the sensor, it aligns with the prototype's focus on functionality and feasibility. Additionally, despite the presence of unused bits in OAP-to-node communication, the efficient utilisation of all 44 bits in node-oriented  communication transmissions optimises the data transmission power consumption for energy-limited nodes. This aligns with the prototype network's energy efficiency goals and achieves a balance between immediate requirements, energy efficiency, and less complex implementation. 

\subsubsection{Energy share parameter (N) optimisation for the prototype network}
Given that this network consists of three DE-LIoT nodes, a less complex energy-sharing parameter optimisation is implemented. For a prototype PSN with a data sampling interval $T_{\text{Int}}$, which includes sensing time (comprising PV, storage, sensor, and data transmission) $t_{\text{Sense}}$ and standby time $t_{\text{Standby}}$, the following equation holds:
\begin{equation}
t_{\text{DataNet}} = t_{\text{Sense}} + t_{\text{Standby}} , \label{eq10}
\end{equation}

Therefore, by substituting the above into (\ref{eq2}), the (\ref{eq5}) can be written as follows:
\begin{equation}
r_{\text{DutyCycle}} = 1 - \Biggl(\frac{t_{\text{DataNetRec}} + t_{\text{Sense}}}{T_{\text{Int}}}\Biggr)
- N\Biggl(\frac{t_{\text{EnergyNetRec}} + t_{\text{EnergyNet}}}{T_{\text{Int}}}\Biggr) , \label{eq11}
\end{equation}

In this prototype network,  MA was based on TDMA. When the data sampling interval is reached, the OAP is expected to request data from each PSN nodes in sequence. After the data request session, the OAP forces all PSNs into ETX sessions to create an energy route, the total duration of which is determined by the calculated \textit{N} value. 
In order to request data from all PSN nodes during the data requesting window $T_{\text{DataReq}}$ and achieve optimum energy sharing in the network, the following condition is required to be met (typically  $t_{\text{DataNetRec}} \ll t_{\text{EnergyNetRec}}$), and therefore, it has a very minimal impact on the network.

\begin{equation}
{t_{EnergyNetRec}}<{T_{DataReq}}  { \, \leq \, }   { t_{Standby}.}  \label{eq12}
\end{equation}
In practice, it is essential to acknowledge that all PSNs may not consistently experience identical illumination levels at any given time. Such variations can arise from minor performance differences in hardware components or fluctuations in channel conditions, impacting the EH performance. In this context, the scenario assumes illumination fluctuations affecting the node's performance, highlighting the need to consider real-world variability in system operation and energy harvesting capabilities. For a fixed $T_{\text{Int}}$, with a node receiving $A$ lux and $A+\delta$ lux illumination, 

$t_{\text{EnergyNetRec}}(A+\delta) < t_{\text{EnergyNetRec}}(A)$,  and due to that from (\ref{eq11}), $t_{\text{Standby}}(A+\delta) > t_{\text{Standby}}(A)$.
Taking into account the above discussion, the following condition can be defined to determine \textit{{T}\textsubscript{DataReq} }for the prototype network.
\begin{equation}
max({t_{EnergyNetRec}}) < {T_{DataReq}} < min({t_{Standby}}) . \label{eq13}
\end{equation}
Based on (\ref{eq13}), with information of maximum \textit{{t}\textsubscript{E:Sleep} }and minimum  \textit{{t}\textsubscript{Standby } }which related to PSN of the prototype network, the optimum\textit{ {T}\textsubscript{DataReq} }for prototype OAP can be determined. In addition, from (\ref{eq11}) and (\ref{eq13}), it becomes apparent that for a configured \textit{{T}\textsubscript{DataReq} }value based on the minimum illumination level for PSNs, nodes experiencing slightly higher illumination conditions can have lower \textit{{t}\textsubscript{EnergyNetRec}}. This suggests that nodes operating above the minimum illumination level have more opportunities to perform ETX compared to PSNs operating at the minimum illumination. In this illumination-based approach, PSN nodes with illumination levels surpassing the minimum value used to calculate \textit{\(T_{\text{DataReq}}\)} immediately increase their \(N\) value after the \textit{\(T_{\text{DataReq}}\)} period within the \(T_{\text{Int}}\) period. This ensures that the nodes in the prototype network achieve an optimal \(N\) value based on the received illumination level. This optimisation is subject to a manually defined minimum \(N\) value, corresponding to the minimum illumination required for PSNs. 

A limitation of this approach is that, for the DE-LIoT network to function effectively, a minimum acceptable illumination level needs to be determined for the environment where the network will be deployed. Although this determination is attainable in less dynamic indoor environments, defining it becomes challenging in highly dynamic settings characterised by fluctuating illumination conditions.

\subsection{Implementation of  LIoT Data- Energy networking enabled OAP}

\subsubsection{OAP architecture}
\begin{figure}[htbp]
\centerline{\includegraphics[width=.65\textwidth]{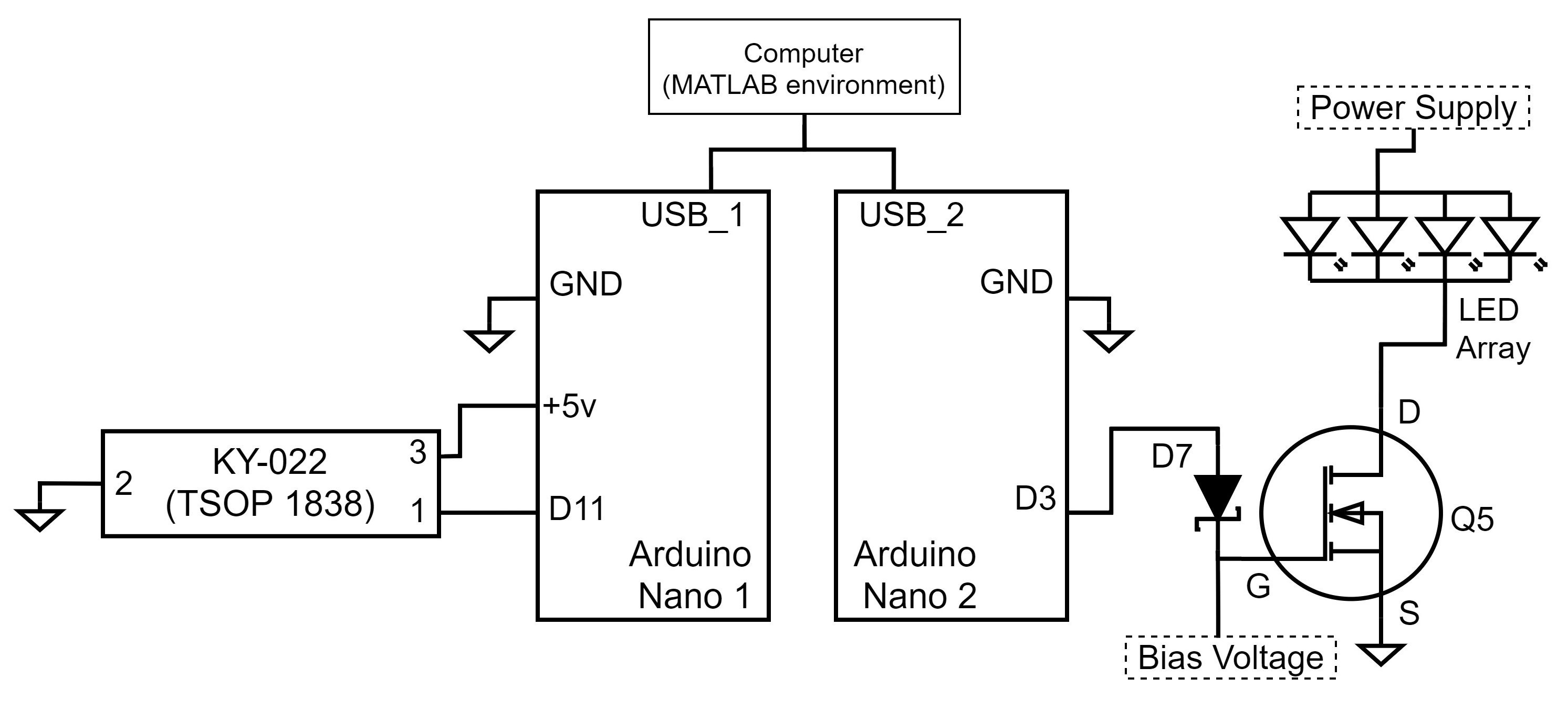}}
\caption{ Circuit diagram of the implemented prototype OAP, with two Arduino Nano boards for simultaneous RX and TX functionalities.}
\label{imp_OAP}
\end{figure}

The prototype optical access point (OAP) was designed to simultaneously facilitate data transmission and reception, while providing eye-friendly illumination sufficient for energising nodes. In order to achieve this functionality while prioritising simplicity, the implementation used two AVR MCU cores. In this design, a computer running the Intel x64 architecture processor with the MATLAB environment was employed to execute all the algorithms. The AVR MCUs were employed as encoders and decoders, as well as interfaces for communication and interaction with other electronic components. Figure \ref{imp_OAP} depicts the implemented OAP circuit diagram.

The LED driver employed an IRF520 MOSFET module to regulate signalling in the LED array. Core 1 of the MCU was implemented using an Atmega328P on an Arduino Nano board, and similarly, Core 2 of the Receiving MCU employed an Atmega328P on an Arduino Nano board. The Receiver integrated a KY-022 module (TSOP 1838) for the detection of incoming signals. The transmitter used a dual white colour LED lamp array (3200K + 6500K), producing a combined spectrum that improves energy harvesting within indoor environments. The LED setup was shielded with a metallic diffuser, allowing for the establishment of distinct illuminated areas for testing. The LED array, with a peak power of 12W, allows for illumination changes on the experiment conducting surface through height adjustment or by altering the power supply input.
  
\subsubsection{OAP algorithm}

\begin{figure}[htbp]
\centerline{\includegraphics[width=.6\textwidth]{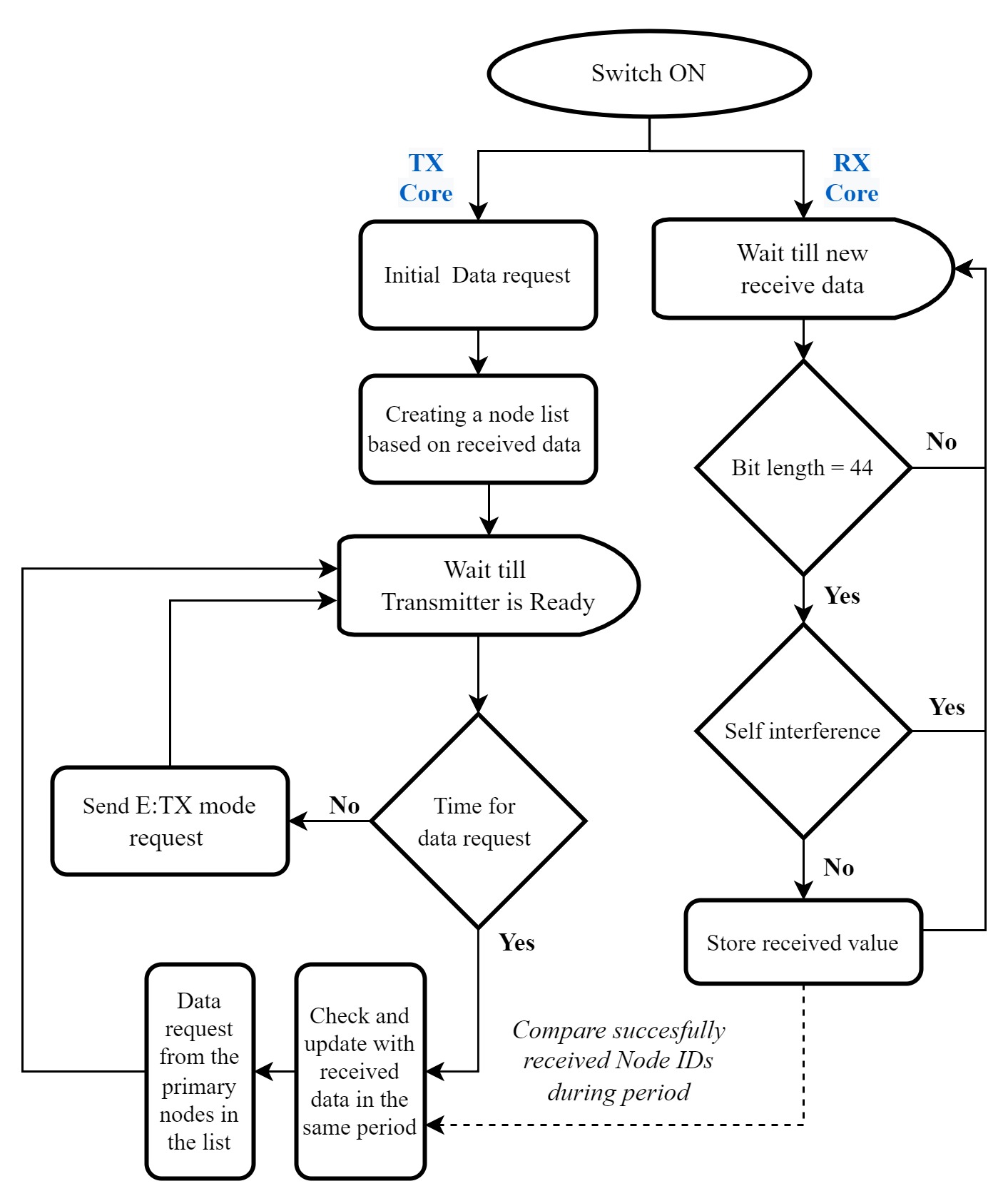}}
\caption{ The algorithm used in the implemented prototype OAP.}
\label{implemented_TX_algo}
\end{figure}

The algorithm is designed to use multi cores in order to execute both transmission (TX) and reception (RX) simultaneously.  For the proof-of-concept prototype, complex optimisation methods to obtain \textit{N} at the OAP were not considered. The algorithm used for the OAP is illustrated in Figure \ref{implemented_TX_algo}. The TX core initiates an initial data request to retrieve important details such as "Node ID," initial voltage levels, and sensor data from local nodes. Subsequently, the OAP continues to send data and ETX requests following the established \textit{{T}\textsubscript{DataReq}} period. In cases where there is not yet time for a data request, the TX core is compelled to transmit ETX requests for the registered PSNs. This approach is implemented to leverage nodes that receive slightly higher illumination levels, allowing them to increase their\textit{ N} values beyond what was initially determined by the OAP based on the minimum illumination requirements for the nodes. Conversely, the RX core algorithm remains in a waiting state for incoming requests. Upon receiving a 44-bit data frame, the RX core distinguishes between nodes and self-interface data, storing values accordingly. In MATLAB, the successful "Node ID" values are compared within the \textit{{T}\textsubscript{DataReq}} interval before reissuing the data request command.

\subsection{Implementation of  LIoT Data- Energy networking enabled Node }
\subsubsection{Node architecture}

\begin{figure}[]
\centerline{\includegraphics[width=.57\textwidth]{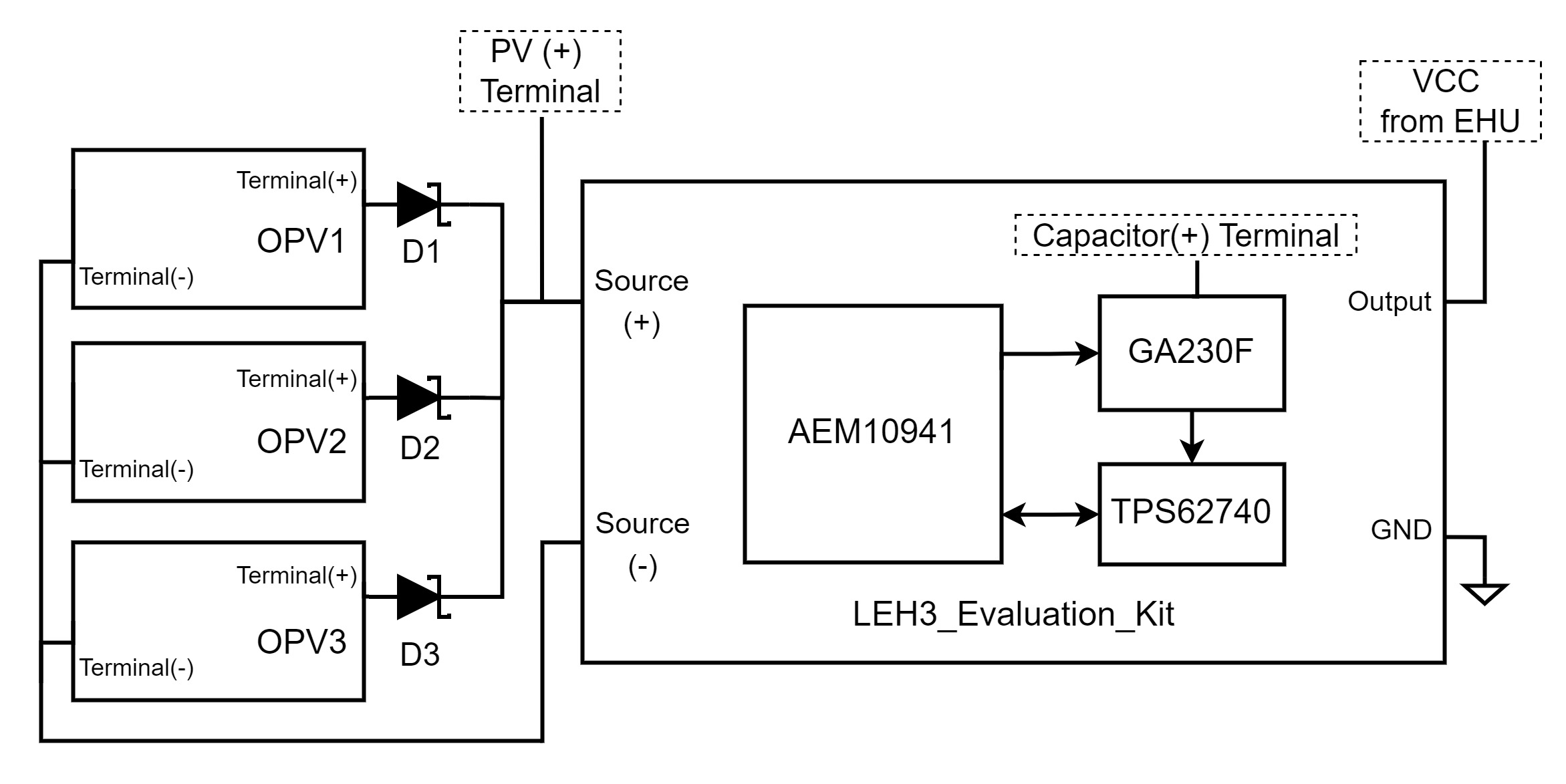}}
\caption{ The diagram of implemented EHU.}
\label{IMP_EHU}
\end{figure}

The schematic of the EHU of the DE-LIoT node is illustrated in Figure \ref{IMP_EHU}. The EHU acquires power using three 50 x 50 mm printed OPV Epshine LEH3 modules interconnected in parallel. This configuration allows efficient energy capture for the node, where each module can generate 0.9 mW of power under 1000 lux illumination \cite{epishine2023leh3}. The PMIC responsible for regulating power is the AEM 10941. Energy storage is achieved through a GA230F 400mF super-capacitor. The DC-DC converter employed is the TPS62740. Since commercially available Epshine LEH3 50 x 50 mm printed organic photovoltaic (OPV) cells have the capability to generate sufficient power under the visible light (VL) spectrum wavelengths, they were selected as harvesters for the node. These OPV cells complement the LW514 white LED, which covers most of the VL spectrum wavelengths, serving as energy transmitters. In the proof-of-concept prototype implementation, angular diversity transmissions and reception scenarios were not considered. Therefore, a single LW514 white LED and PD were used to perform ETX and detect signals from a single direction.

These PV cells are strategically oriented within the node casing to efficiently capture energy from diverse directions. Schottky diodes (D1, D2, D3) are used at the positive terminals of the PV cells to prevent reverse voltage problems. For the remaining energy harvesting components, the Epshine LEH3 evaluation kit is used, integrating an AEM 10941 PMIC, GA230F for energy storage, and the TPS62740 DC-DC converter. The AEM 10941 PMIC's MPPT configuration is preset by the manufacturer to match Epshine LEH3 PV cell requirements. 
The Epshine LEH3 evaluation kit employs the AEM 10941 PMIC with specific voltage configurations, such as a minimum storage voltage for cutoff ({\textit{V}}\textsubscript{\textit{OVDIS}}) set at 3.2V and a threshold for output re-enable ({\textit{V}}\textsubscript{\textit{CHRDY}}) at 3.8V. It is critical to prevent the voltage from dropping below the {\textit{V}}\textsubscript{\textit{OVDIS}} threshold to avoid complete power termination until {\textit{V}}\textsubscript{\textit{CHRDY}} is achieved, thereby minimising the risk of data loss. Wire connections to test points are established to monitor capacitor and PV terminal voltages, mitigating the potential for nodes to become unresponsive. Fine-tuning the DC-DC converter to provide a 3.2V output optimises node power consumption and ensures accurate functionality of voltage-dependent features, including the ADC. 

\begin{figure}
\centerline{\includegraphics[width=.45\textwidth]{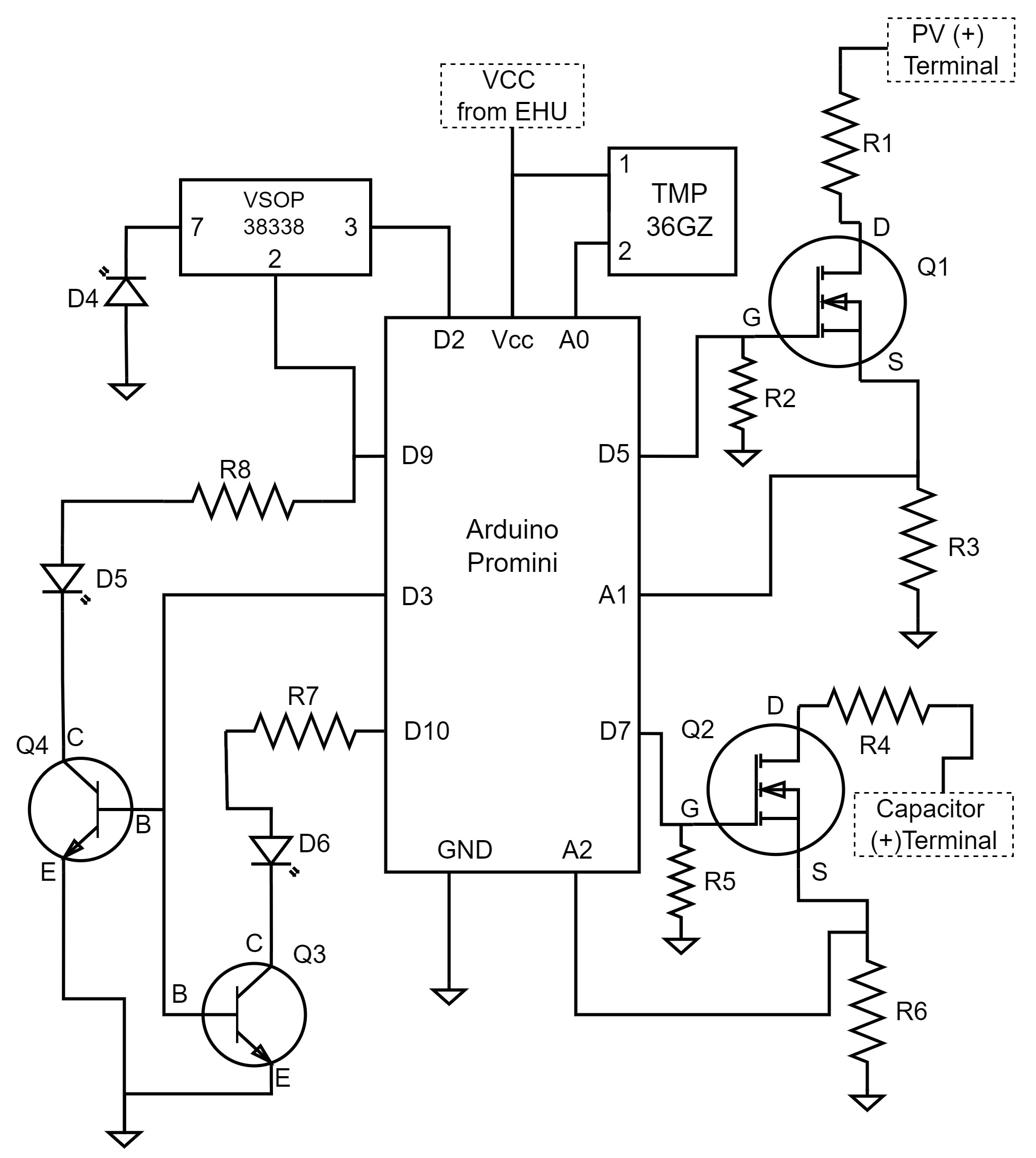}}
\caption{ The circuit diagram of implemented DE-LIoT node.}
\label{DE_LIOT_CIRCUT}
\end{figure}

Figure \ref{DE_LIOT_CIRCUT} presents an overview of the setup, illustrating the MCU, the transceiver, and related circuitry components. The MCU driving the node is an Atmega328P, operating at 3.3V and 8MHz. The optical detector uses a silicon 525 nm photo-diode for precise light sensing. The trans-impedance amplifier (TIA) and pre-amplifier incorporate the VSOP38338. For data uplink, an IR 3 mm LED is utilised, while the Energy/Data transmitter adopts the LW514 LED, featuring a half-intensity angle of 15 degrees.

Since the DC-DC converter of the EHU ensures a stable voltage, and to further minimise power consumption for the MCU, the inbuilt voltage regulator and power-on LED of the Arduino Pro Mini 3.3V 8MHz development board were removed. N-channel power metal-oxide-semiconductor field-effect transistor (MOSFET)s Q1 and Q2 were used as switches to isolate the capacitor and PV voltage measurement sub-circuits, assisting energy harvesting performance. These MOSFETs, activated by high signals from GPIO pins 9 and 11, establish connections between resistors R1, R3, R4, and R6, forming a voltage divider setup. This division reduces PV cell and capacitor voltages for safer ADC measurement via pins 1 and 2. After measurement, the MCU sets GPIO pins 9 and 11 to a low state, disconnecting the divider circuit from the EHU and maintaining the energy harvesting process undisturbed. The transceiver's amplifier circuit employed the VSOP38338 IC, which integrates essential functions such as demodulation and decoding. It encompasses an automatic gain control amplifier, a band-pass filter, and a comparator to transform PD D4's current waveform into a voltage signal. This signal is then separated into useful data, converted to a digital waveform, and transmitted to the MCU through the VSOP38338 IC. For communication with the access point, a 3mm IR LED (D5) was used, controlled by Q4. The energy transmission and communication between nodes involved a D6 LW514 LED and a Q3 transistor. 

The power to the transceiver circuits was supplied through GPIO pins 13 and 14, allowing their activation or deactivation to reduce power consumption. A single GPIO pin (GPIO 1) facilitated modulated data output across all transmitters, with the algorithm selectively powering the chosen transmitter via GPIO pins 13 and 14. The node used the TMP36GT9Z sensor as sensing device. The output of TMP36GT9Z was linked to ADC0 for measurement purposes.

\subsubsection{Node algorithm}
  
\begin{figure}
\centerline{\includegraphics[width=.68\textwidth]{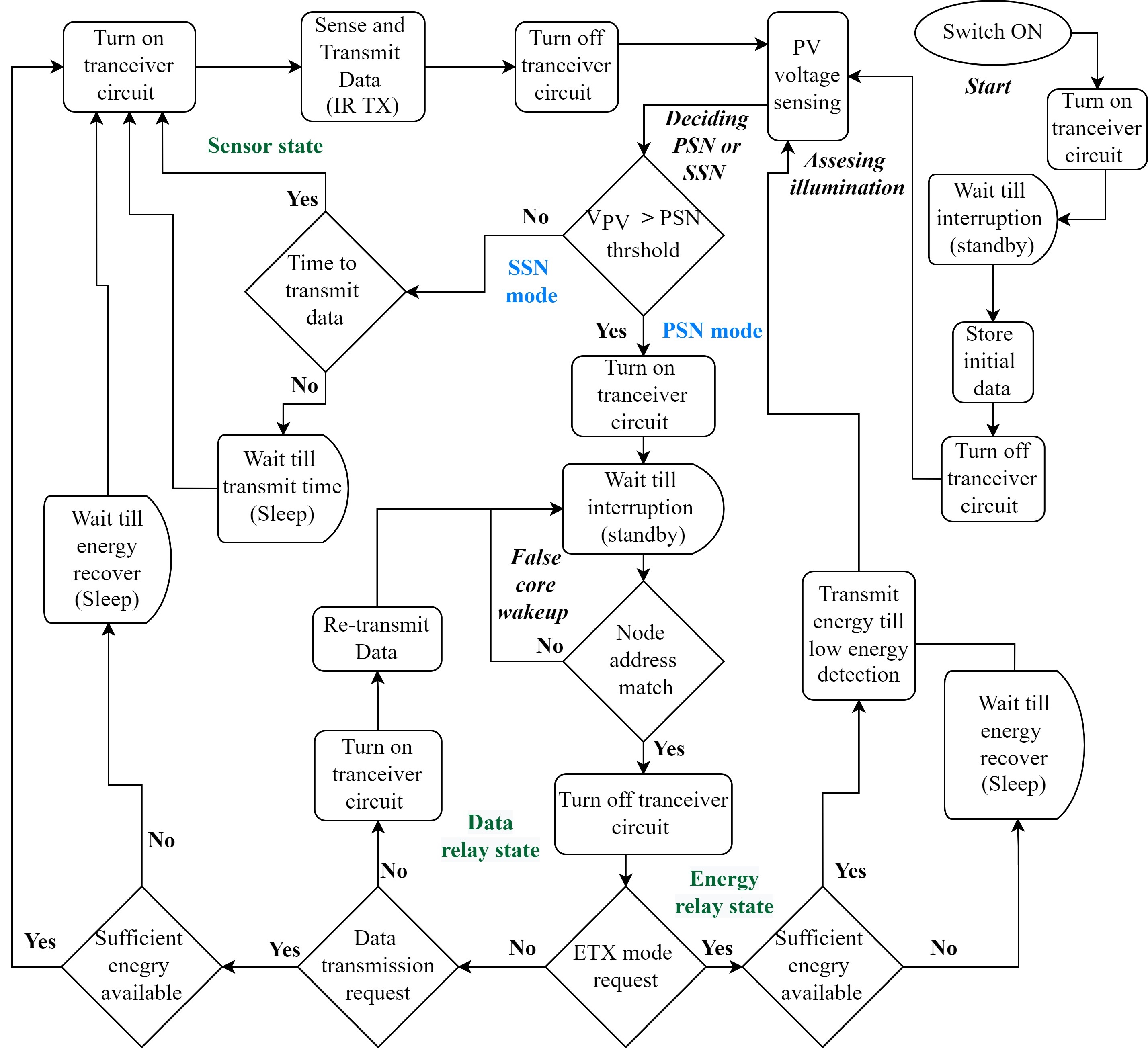}}
\caption{ The simplified algorithm of the implemented prototype DE-LIoT node. }
\label{node_ALGO}
\end{figure}
The algorithm used for the execution of tasks in the prototype nodes is illustrated in Figure \ref{node_ALGO}. The node algorithm is designed to follow soft intermittent operation in order to minimise node outages. The node algorithm comprises PSN and SSN modes, with the decision of the role being determined based on available resources such as PV illumination. Furthermore, algorithm encompasses three node states: the data relay state for inter-node communication and the energy relay state for energy sharing, both of which are activated only when the node is in PSN mode. The sensor state, which senses and transmits sensor data, is common for both PSN and SSN, as the main objective of the sensor nodes is to transmit temperature data to the OAP. 
Prototype algorithm expects nodes to establish better LOS communication with the OAP at the start. Initially, the nodes are designed to be receptive to information from the OAP. In order to facilitate this, the node engages the transceiver circuit and concurrently disables the core, initiating a state of "standby mode". When a signal is detected, the node reactivates the core and initiates the process of decoding the received message. Through this procedure, nodes collect and retain initial data, such as \textit{{T}\textsubscript{Int}}. As per the algorithm, the node is programmed to autonomously select its operating mode based on the measurement of the PV terminal voltage (\textit{{V}\textsubscript{PV}}). In order to mitigate sudden spikes in PV terminal voltage caused by the 82 ms window for open voltage value measurement during MPPT performed by the AEM 10941 PMIC, three distinct voltage readings are collected. These readings are taken at 50 ms intervals, and the minimum value among them is selected as\textit{ {V}\textsubscript{PV}}.   For simplicity, this prototype network implementation assumes a static channel environment. Consequently, the PSN or SSN decision process relies solely on a single \textit{{V}\textsubscript{PV}} threshold comparison. Therefore, the nodes were classified as PSN if their \textit{{V}\textsubscript{PV}} $>$ 3.0V. In the presence of dynamic channel conditions, the approach could incorporate several readings over time to minimise erroneous decision-making.

In the PSN mode, these nodes remain idle, listening to incoming requests from either the OAP or neighbouring nodes, and responding as necessary. Upon receiving an interruption, the node wakes up the MCU core from its standby state to decode the received message. Initially, since a correlator was not used in the design, the system verified whether there was a match with the designated address before proceeding to interpret the instructions within the payload. In the event of an address mismatch, the node recognised this as a false wake-up call. As a result, it deactivates the core once more and reverts back to standby mode, with the intention of minimising power consumption. 
In this work, the "Node ID" is used to differentiate between the OAP and other nodes. A matching "Sender ID" indicates an OAP data frame, whereas different IDs signify data-relay state requests. In this prototype work, while in the data-relay state, the PSN directly forwards received data to OAP while preserving the origin "Node ID". 
On the other hand, for frames from the OAP, the "command ID" directs subsequent actions. These could involve entering energy relay state or sensor state. The node follows the instruction given by the "Command ID". In general, the developed algorithm checks the energy sufficiency for the assigned task before execution by measuring the storage energy availability. As a trade-off, this approach increases power consumption for each state. However, it ensures that the node will not experience power outage during any state. If insufficient energy is detected for the task, the algorithm forces the PSN into a sleep mode until sufficient energy becomes available. For implementation, this was achieved by measuring the internal voltage of the capacitor and comparing it with a predefined threshold level. During energy-relay state, the node activates ETX operation and monitors the voltage of the capacitor (\textit{{V}\textsubscript{Capacitor}}), stopping ETX when\textit{ {V}\textsubscript{Capacitor} }reaches approx \textit{{V}\textsubscript{min}}.  If \textit{{V}\textsubscript{PV}} $<$ 3.0, nodes switch to SSN mode, activating the transceivers only in the transmission windows to minimise power use. According to\textit{ {T}\textsubscript{Int }} value received in the initial phase, the SSN wakes up using the internal timer and transmits sensor data. The end of each cycle shows a self-assessment of the PV terminal voltage and recent illumination levels, determining the operating mode. 

\begin{figure}
    \centering
    \includegraphics[width=0.67\linewidth]{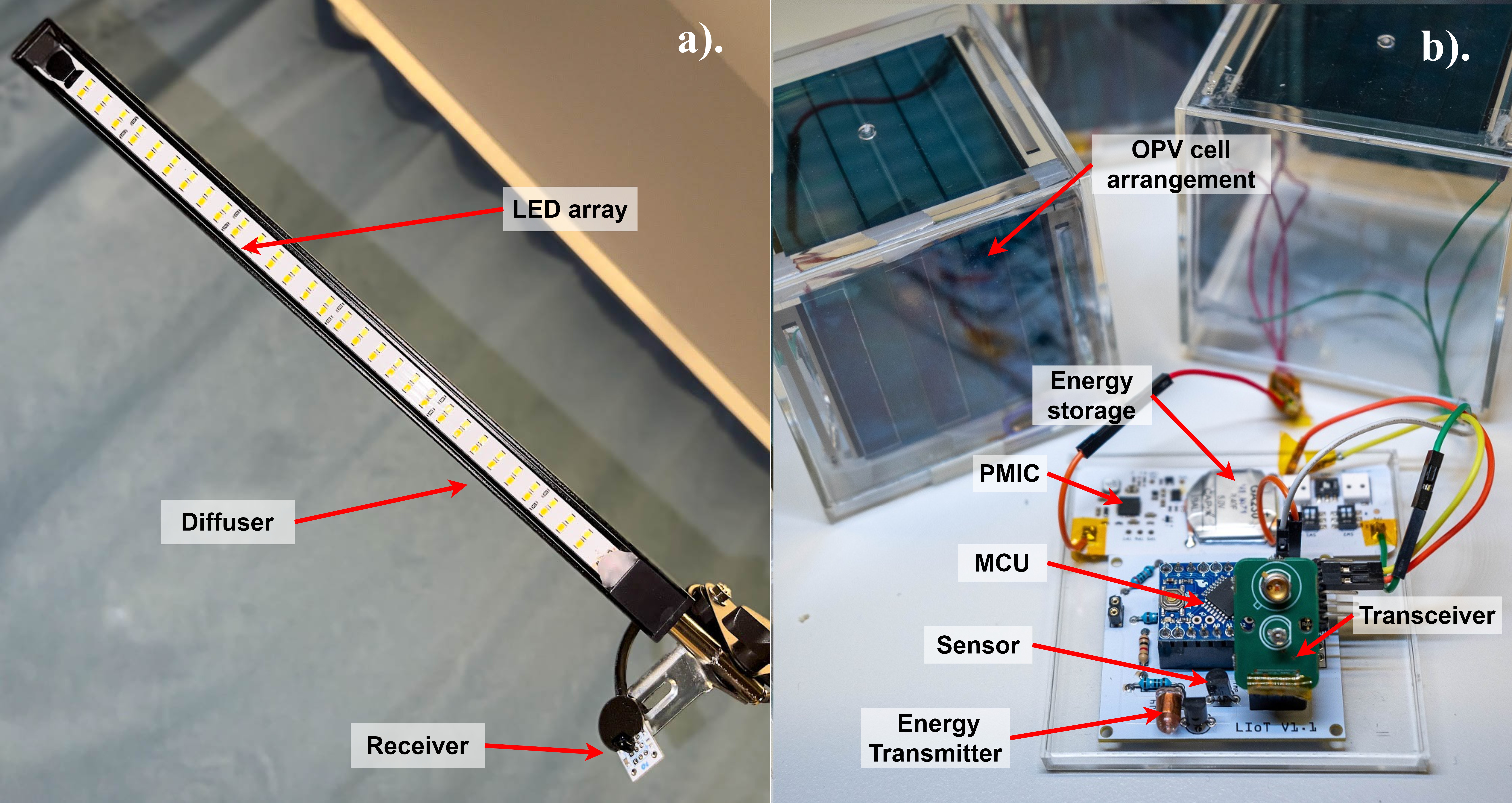}
    \caption{View of the implemented DE-LIoT prototypes: a) OAP transmitter LED array and receriver; b) Nodes.}
    \label{node_photo}
\end{figure}

In the final configuration, each node was encased within a transparent cubic enclosure, resulting in physical dimensions of 7.3 x 7.3 x 7.3 cm for each unit.  The finalised DE-LIoT nodes and OAP proof of concept prototypes for the experiments are shown in Figure \ref{node_photo}. 

\section{Performance evaluation}

In this section, the evaluation of the prototype DE-LIoT nodes and network, developed based on the information described in the previous section, is carried out. In experiments, a minimum illumination level of 1000 lux is employed to ensure that PV cells and EHUs operate at their peak performance within PSNs, aligning with the defined range of indoor lighting standards.

\subsection{Characteristics of the prototype DE-LIoT node}

\begin{figure}[]

\centerline{\includegraphics[width=.48\textwidth]{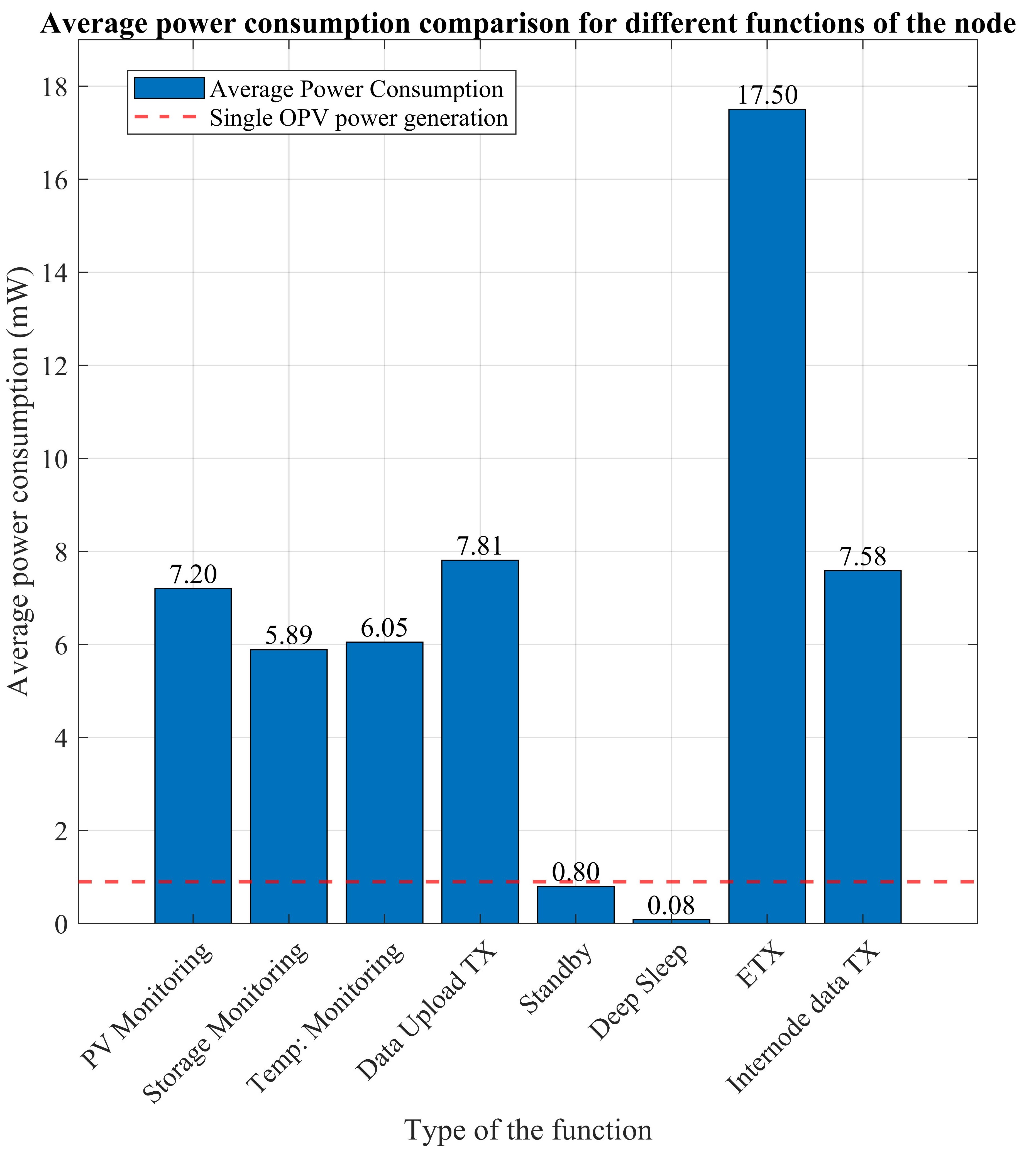}}
\caption{ Average power consumption of the functions compared to the power generation of a single OPV cell under 1000 lx illumination.}
\label{powerprofile}
\end{figure}

In this evaluation, the power consumption for each functionality described in previous section is measured. In order to obtain power consumption data related to various functions of the DE-LIoT node, the Nordic Power Profiler Kit II was employed. The results obtained are illustrated in Figure \ref{powerprofile}.
Based on the measured results, the node experiences maximum power consumption during ETX and minimal consumption during sleep mode. It can be observed that the power consumption of standby mode, which is vital for data relay mode, and sleep mode, which is vital for energy relay mode to rapidly generate surplus energy, is below the power generation of a single OPV cell used for this prototype under 1000 lx illumination. All other functions exhibit similar power consumption levels. Therefore, the component selection, design, and algorithm approaches on the prototype nodes were able to operate the nodes in data and energy relay modes with surplus energy generation, preserving energy autonomous operation, which met the expectations of the DE-LIoT concept node prototype. 

It is crucial that DE-LIoT nodes optimise their power consumption in varying functional modes, with a primary focus on minimising energy consumption without compromising data and energy transmissions. Functions in which nodes predominantly operate, particularly standby and deep sleep, are crucial, enabling the conservation of surplus energy for critical functions such as inter-node data and energy exchange. This strategy enhances the establishment of extended data energy routes, ensuring prolonged network longevity. This benefits nodes in scenarios with limited harvestable energy, especially in dynamic indoor environments, allowing them to operate longer.

\subsubsection{Energy transmission as interference}
\begin{figure}[htbp]
\centerline{\includegraphics[width=.52\textwidth]{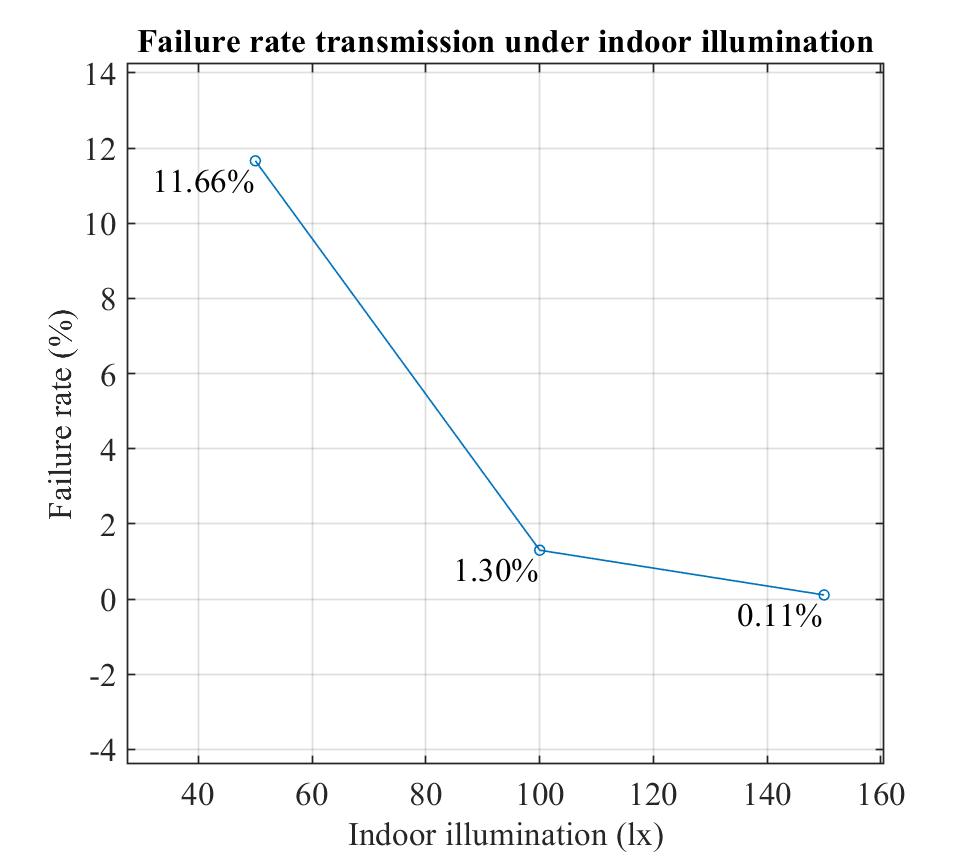}}
\caption{ The impact of node-node ETX  for OAP-node communication link.}
\label{failure}
\end{figure}

In DE-LIoT systems, where both data and energy networks operate within the same spectrum range, the energy networking aspect has the potential to introduce interference to the data network links. Therefore, the following experiment was conducted to determine how energy transmission interferes with SSNs. Since the signal strength received from the OAP is analogous to the illumination level, the prototype node was placed in an area with minimal illumination approx 150 lx.  For the accurate measurement of illumination, a Chauvin Arnoux 111C lux meter with an LED source measuring configuration was used. Additionally, a PSN is configured to operate using the ETX mode. Same time OAP is used to transmit data to the SSN. Then the failure rate of the data frames was calculated. The experiment was repeated by increasing the illumination level on the SSN. The observed variation is graphed in Figure \ref{failure}.

The graph in Figure \ref{failure} illustrates that placing the SSN in an area with low signal strength (i.e., low illumination) in an energy transmission scenario can have a notable impact on the communication performance between the OAP and the node. Since the system operates within the same spectrum region for both data and energy networking, the optical detectors may become saturated by energy transmission bursts. As a result, these detectors might fail to detect weak signals originating from the OAP. Therefore, SSNs positioned far away from the OAP, where low illumination is more prevalent, are more prone to experiencing a higher transmission failure rate compared to SSNs with better illumination. By adaptively adjusting the assignment of PSN and SSN modes based on actual illumination, the DE-LIoT concept inherently minimises low-light scenarios, thereby reducing network failure rates. Furthermore, to reduce the transmission failure rate, the following approaches can be introduced. For the same spectrum data energy networking in DE-LIoT systems, strategies such as employing multiple data transmissions for SSNs and scheduling between energy and data networking, depending on the application, can be used to enhance the quality of service and mitigate the impact of ETX as interference. On the other hand, opting for distinct spectra for OAP and inter-node data-energy networking and implementing physical wavelength filtering can effectively segregate data signals, particularly under low signal strength scenarios attributable to spectrum differences.

\subsection{Network performance evaluation}

\begin{figure}[]
    \centering
    \includegraphics[width=0.5\linewidth]{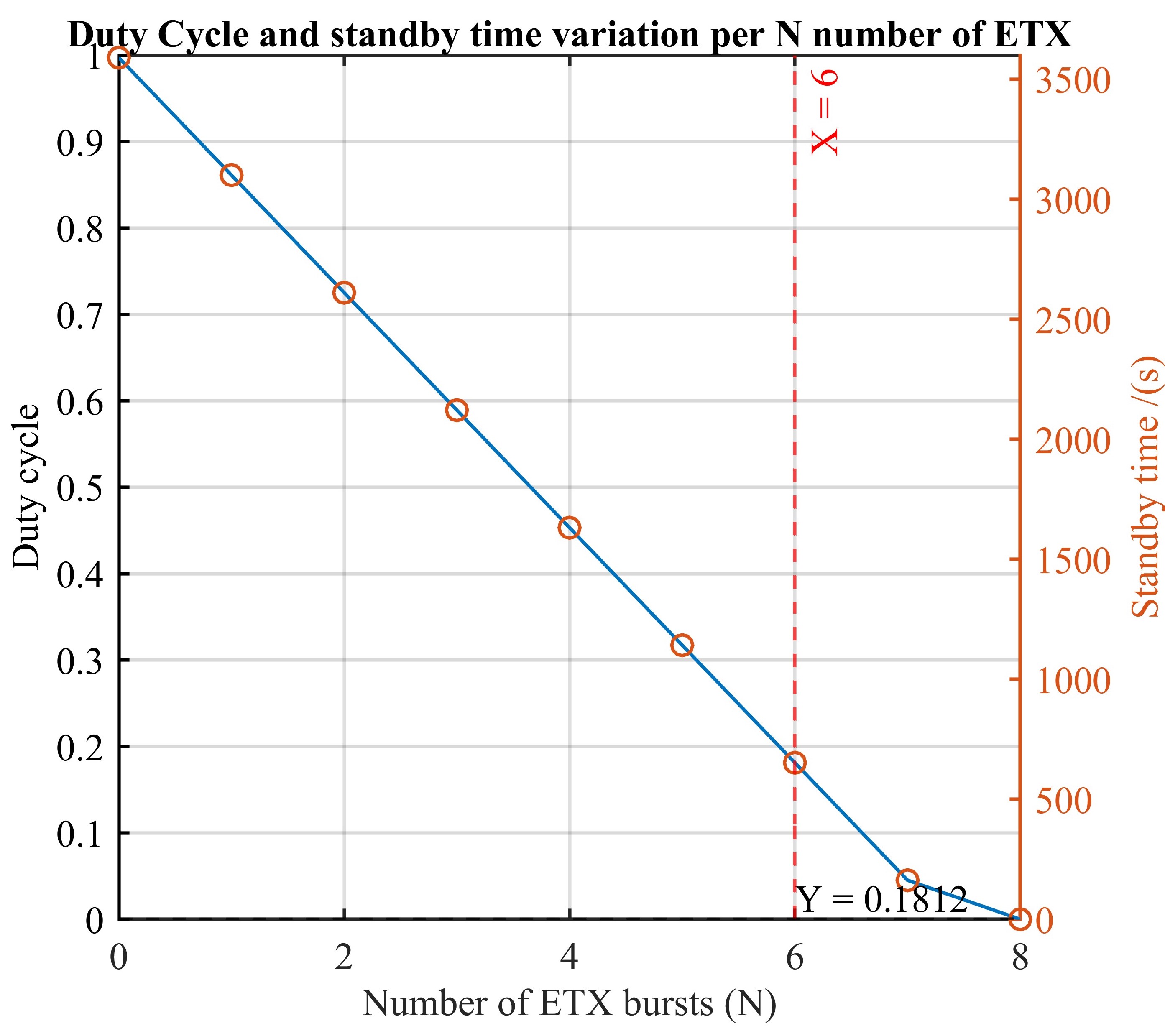}
    \caption{Variation of duty cycle and standby time with \textit{N} value for the prototype DE-LIoT node under 1000 lx illumination. }
    \label{Nval}
\end{figure}

Throughout the experiments, the VLC channel remained static, with the expectation that the PSNs would receive a minimum illumination of 1000 lux. The measured time parameters for a prototype DE-LIoT node are detailed as follows. \(t_{\text{Sense}}\) is 9.53 seconds, \(t_{\text{EnergyNet}}\) is 40 seconds, \(\textit{max}(t_{\text{EnergyNetRec}})\) is approximately 450 seconds, \(t_{\text{DataNetRec}}\) is approximately 40 seconds. \(T_{\text{Int}}\) was selected as 3600 seconds, aligning with the requirements of the hourly temperature measuring sensor system. In order to measure the impact of the energy networking function on the prototype network, all PSNs were required to operate with higher \textit{N} values, which prioritise the ETX operation, effectively causing them to function as energy relays. The variation of the duty cycle with different \textit{N} values of the prototype under 1000 lx can be visualised using (\ref{eq11}) as illustrated in Figure \ref{Nval}.

Based on Fig. \ref{Nval},  the \textit{N} value for the network is selected as 6 with duty ratio of 0.18 for all PSN nodes. In addition,based on standby time for \textit{N}=6,  \textit{{T}\textsubscript{DataReq} }was selected as 600s. 

\subsubsection{ETX operation performance evaluation}
\begin{figure}
\centerline{\includegraphics[width=.52\textwidth]{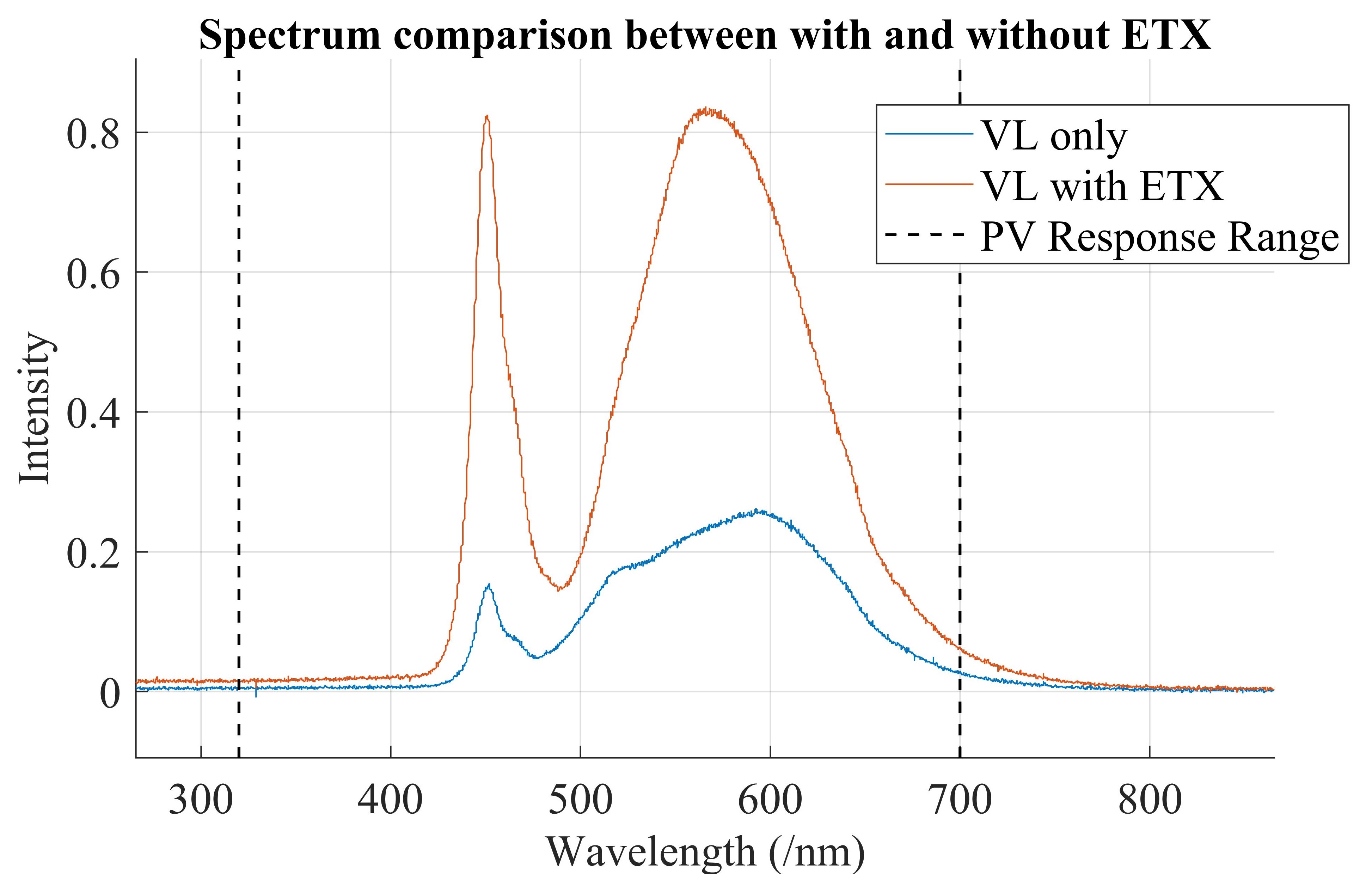}}
\caption{ Spectrum comparison under OAP prototype's VL illumination and VL based ETX from DE-LIoT node.}
\label{spectrum_VL}
\end{figure}

\begin{figure}[H]
\centerline{\includegraphics[width=.57\textwidth]{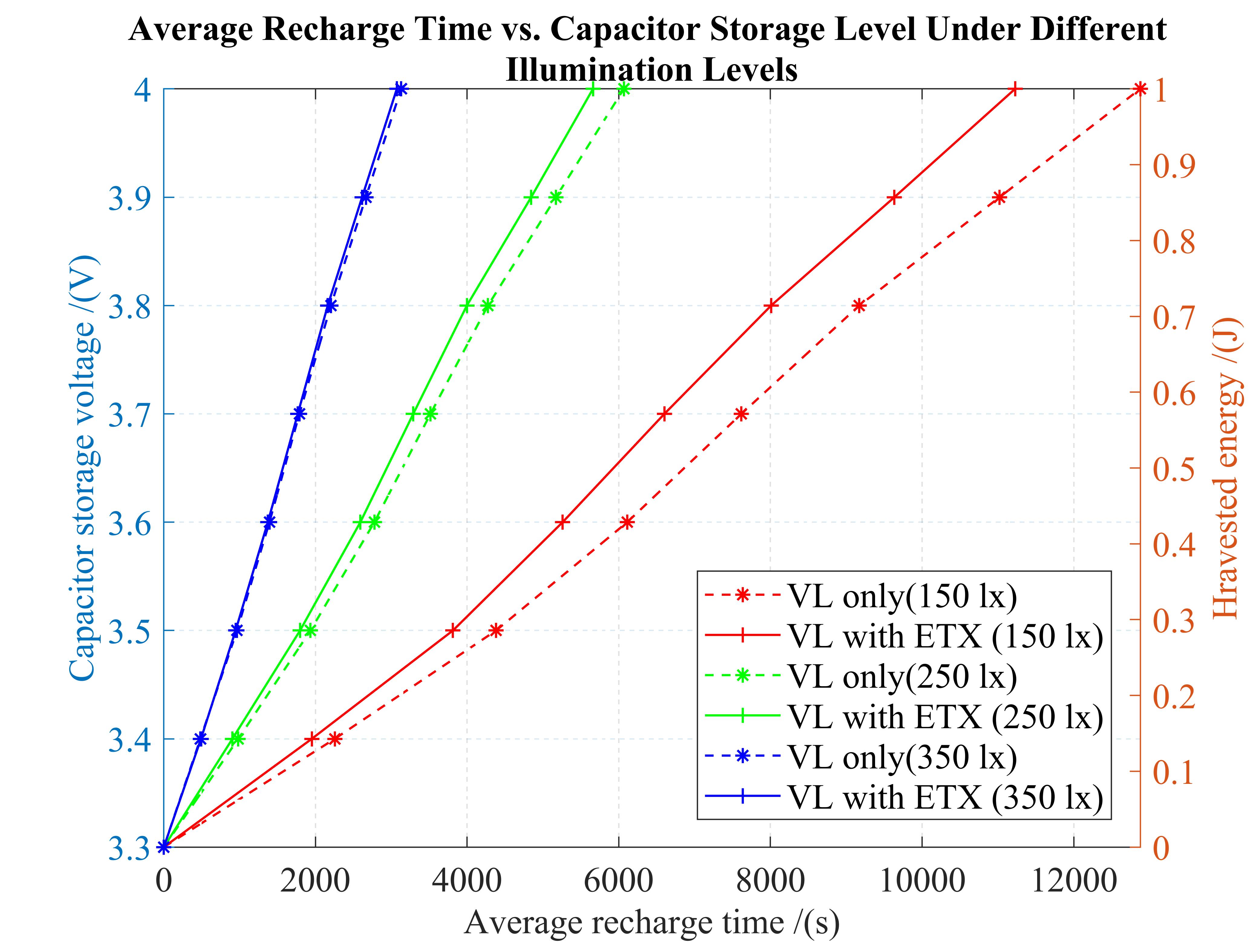}}
\caption{ 
Variation in average recharge time for harvesting 1000 mJ of excess energy across different illumination levels.}
\label{excess}
\end{figure}

For this experiment, two DE-LIoT PSNs and one node were configured to operate in deep sleep mode, representing the minimum power consumption operation mode while maintaining memory functionality. Then node was then placed 20 cm away from PSNs in a low-illuminated area to simulate SSN conditions. The ETX process with PSN nodes was then started.  Afterwards, the experiment was repeated with an increase in the illumination level.  The intensity distribution of the VL spectrum during ETX was observed using a Thorlabs CCS200 spectrometer with CCSB1 cosine corrector. The spectrometer utilises  cosine corrector to receive the free-space optical wavelength spectrum and evaluate optical intensity per wavelength at the SSN location. The values were acquired using Thorspectra software, incorporating amplitude correction with a factory-calibrated configuration to ensure precise comparison of wavelengths. The PV response range is directly sourced from the manufacturer's data sheet. The obtained spectrum intensity variation is graphed in Figure \ref{spectrum_VL}.
The graph clearly demonstrates that the OAP illumination effectively covered the majority of the response range of the Epshine LEH3 PV cell. As the intensity of wavelengths within the same range increases, it is reasonable to anticipate a higher energy harvesting rate when ETX is enabled in the setup. Following this, the \textit{{V}\textsubscript{Capacitor} }of the SSN was observed using a PicoScope 3403D digital oscilloscope and Picoview monitoring software. In the experiment, the average time taken to harvest 1000 mJ of excess energy while the node operated in minimum power mode was evaluated. The observed variations in average recharge time versus capacitor energy level are illustrated in Figure \ref{excess}.
According to the results, a 12.8\% improvement in recharge time is observed when the SSN operates under 150 lx. When the node is exposed to 250 lx, a 6.7\% improvement can be observed. It is also evident that the enhancement from energy transmission decreases as indoor illumination increases. 
As a result, with improved indoor VLC channel conditions for the SSN, the OAP becomes aware of the reduced impact of ETX on the SSN. Consequently, it has the freedom to reduce \textit{N }and increase the duty cycle/standby time to enhance data communication. On the other hand, during night time or when no humans are present in the indoor environment, OAP can reduce room illumination and save energy, increasing the \textit{N} value to compensate for the degraded VLC channel condition due to reduced illumination. 

\subsubsection{Effect of ETX in DE-LIoT sensor network}
\begin{figure}
    \centering
    \includegraphics[width=.3\textwidth]{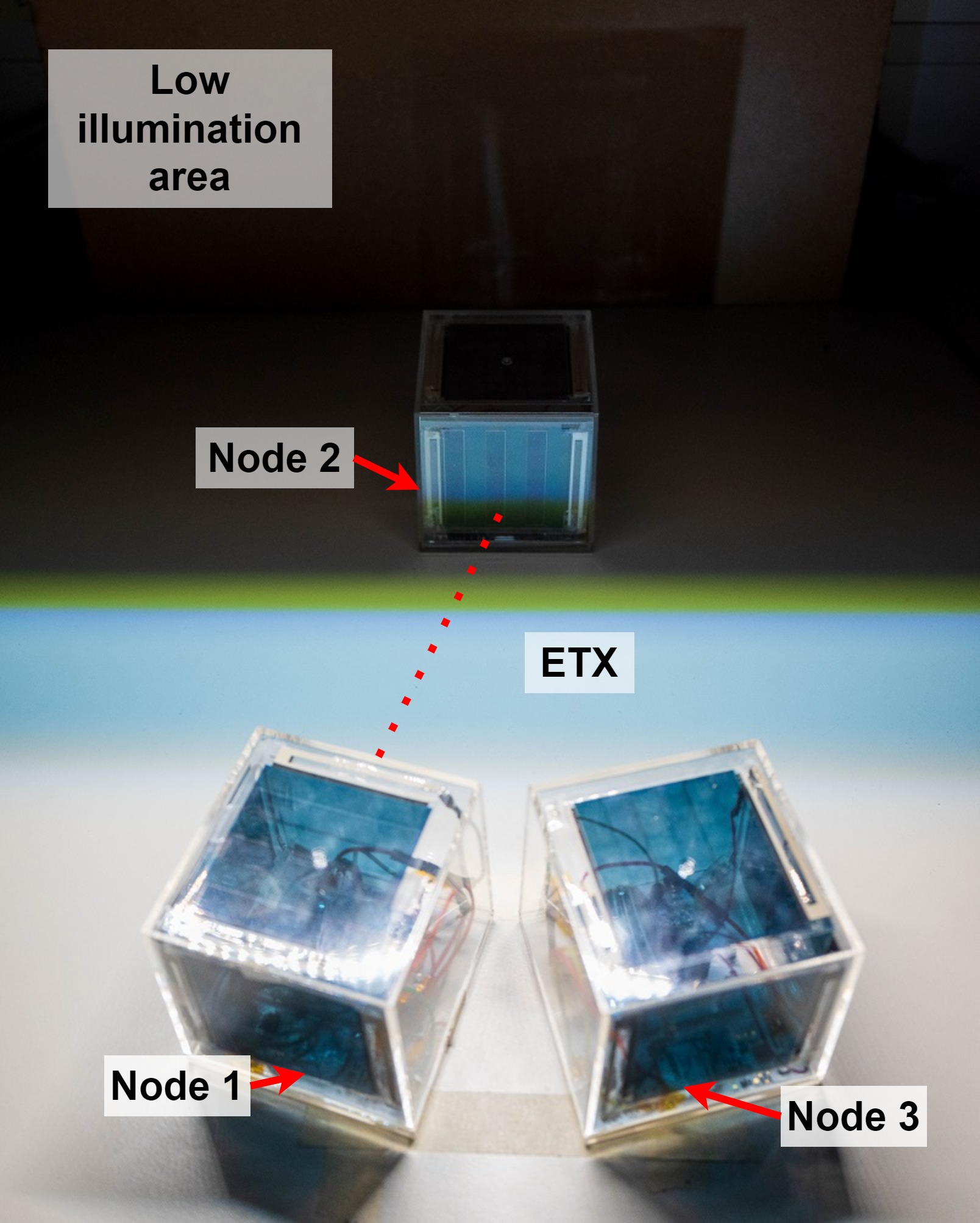}
    \caption{The positioning of prototype PSNs represented by Node 1 and Node 3 (PSNs), along with the SSN represented by Node 2 (SSN), is shown under the illumination provided by the prototype OAP. In order to establish distinct illumination regions, an LED lamp of OAP equipped with a concentrator was employed.}
    \label{long_exp}
\end{figure}

In this experiment, two PSNs (Node 1 and Node 3) and one SSN (Node 2) were employed. Initially, these nodes were positioned 15 cm apart, and their energy storage units were fully charged. The two PSNs were strategically oriented so that their energy-transmitting LEDs were facing the PV face of the SSN. The OAP employed a local-type illumination approach, positioned approximately 40 cm above the nodes' PV surfaces, to establish well-defined low and high illumination scenarios for the experiment. The SSN was placed under approximately 150 lux of nominal illumination on a PV face, while the PSN nodes were exposed to a minimum approx 1000 lux of illumination. The arrangement of the nodes is illustrated in Figure \ref{long_exp}.

\begin{figure}
\centering
\includegraphics[width=1\linewidth]{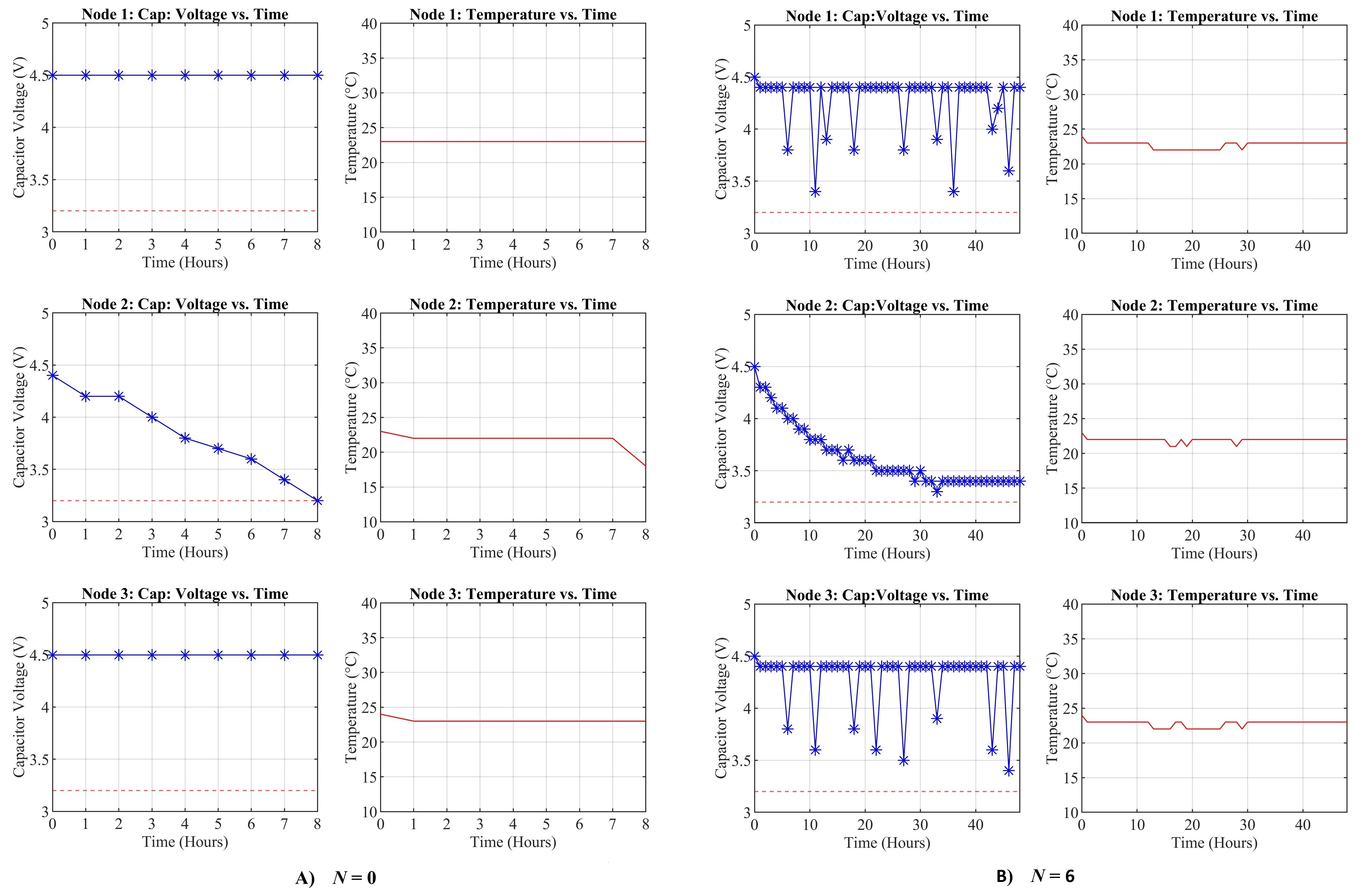}
\caption{The received capacitor voltages and sensed values from prototype DE-LIoT nodes are presented for scenarios without ETX (\textit{N} = 0) and with ETX (\textit{N} = 6).}
\label{No_ETX}
\end{figure}

In the initial phase of the experiment, the system operated without ETX (\textit{N = 0}). The PSNs remained active continuously, without employing sleep mode or engaging in ETX. In contrast, the SSN remained in sleep mode until its internal clock synchronised with \textit{{T}\textsubscript{Int}}. Subsequently, the SSN would proceed to sense and transmit data through one-way asynchronous communication. During each \textit{{T}\textsubscript{DataReq}} window, all PSNs were instructed to transmit capacitor voltage, PV voltage, and the sensing values, allowing for bidirectional synchronous communication among both PSNs. In the latter part of the experiment, the PSNs were directed to operate with ETX (\textit{N = 6}). At the beginning of each experiment,\textit{ {V}\textsubscript{Capacitor}} was maintained at the maximum level. The obtained results are presented in Figure \ref{No_ETX}.

While Figure \ref{No_ETX} illustrates the data transmitted from nodes and received by the OAP, including capacitor voltage and sensed temperature variations, the transmitted and received PV voltages for both PSNs and the SSN remained constant, as the channel conditions were static throughout the duration of the experiment. As shown in Figure \ref{No_ETX}a , in the absence of ETX, the SSN could only sustain operation for 8 hours. Notably, the \textit{{V}\textsubscript{Capacitor} }exhibited a decline during each transmission cycle, and the harvested energy proved insufficient to meet the average power consumption requirements of the SSN node in this specific scenario. Throughout this period, all other PSNs maintained a consistent\textit{ {V}\textsubscript{Capacitor}} level without any fluctuations. This implies that the PSNs were able to maintain a constant voltage by operating as data relays (\textit{N = 0}) throughout the time but did not contribute additional energy generation toward energy networking purposes. Consequently, the harvesting resources of the PSNs were not utilised optimally during the duration of the experiment, and they were solely employed for data networking purpose. As illustrated in Figure \ref{No_ETX}b, it becomes evident that, due to the additional energy acquired from the ETX function, the SSN node successfully accumulates sufficient energy to sustain its operation throughout the 48-hour operational period under consideration. Furthermore, it is noticeable that the \textit{{V}\textsubscript{Capacitor} }of the PSNs fluctuates due to the ETX processes. This signifies that the energy harvesting resources of the PSN nodes actively contribute to the networking by serving as energy relays for the SSN.

\begin{figure}
\centerline{\includegraphics[width=.47\textwidth]{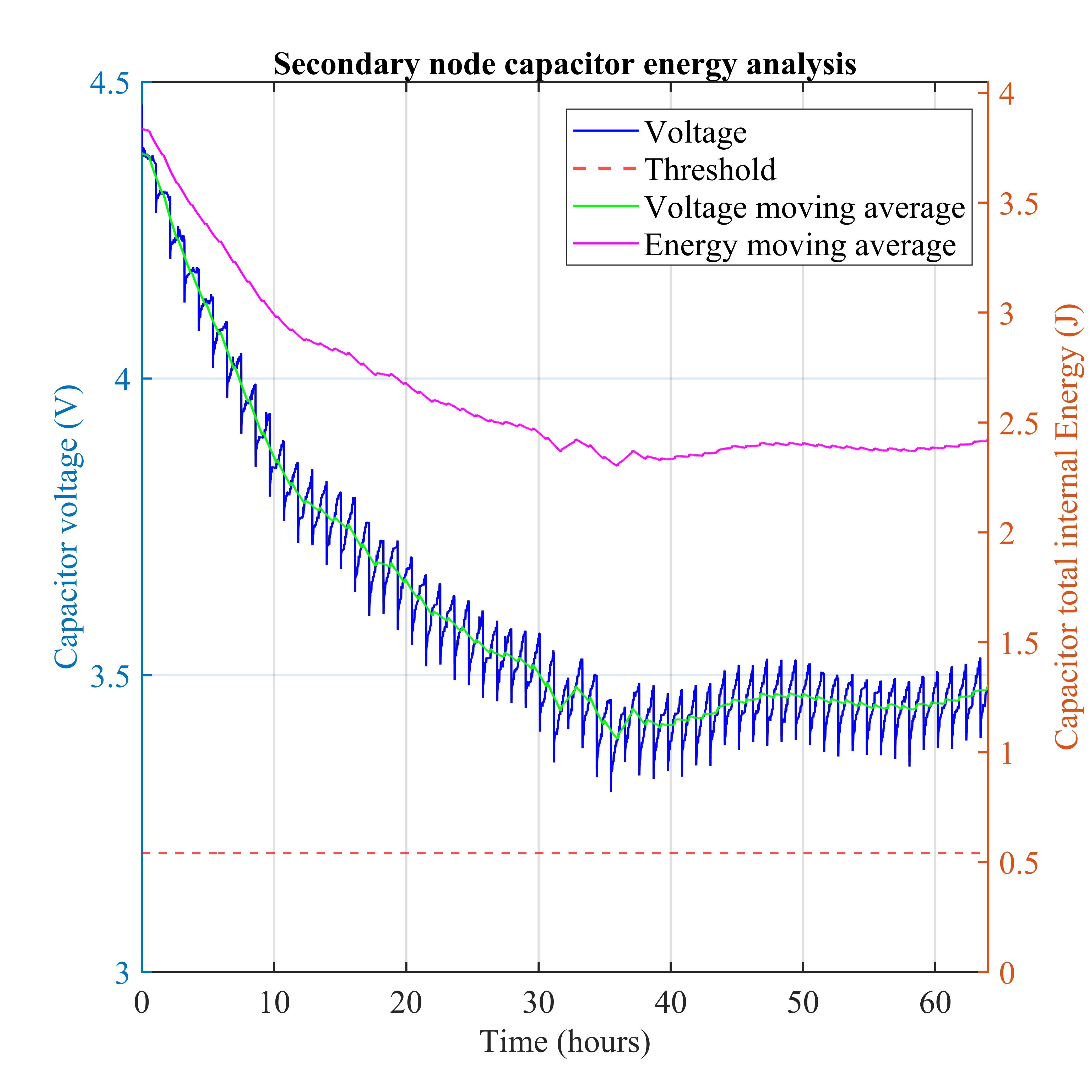}}
\caption{ The variation in SSN \textit{V\textsubscript{Cap}} and total internal energy during the enabled ETX mode over a 60-hour interval. 3.2V (\textit{{V}\textsubscript{OVDIS}}) is considered as the minimum voltage threshold to determine ES depletion.}
\label{SSN_CAP}
\end{figure}

In order to further investigate the recharging behaviour of the SSN, the experiment was repeated while monitoring the probe connection to the internal super-capacitor of the SSN. The voltage measurements obtained from the SSN during 60 hours of operation are presented in Figure \ref{SSN_CAP}.

As shown in Figure \ref{SSN_CAP}, it can be observed that \textit{{V}\textsubscript{Capacitor} }gradually decreases from 4.5V to approximately 3.5V during the initial 20 hours. Subsequently, it remains stable at around 3.5V for the duration of the operation. This behaviour can be attributed to the non-linear characteristics of capacitor charging. Capacitors tend to accumulate charges more rapidly when voltage levels are low and slow down as the voltage approaches its maximum level. Furthermore, from the energy moving average curve, it becomes apparent that the total energy of the SSN node converges to a steady value as the 60-hour experiment time elapses.

\begin{figure}[H]
\centerline{\includegraphics[width=.67\textwidth]{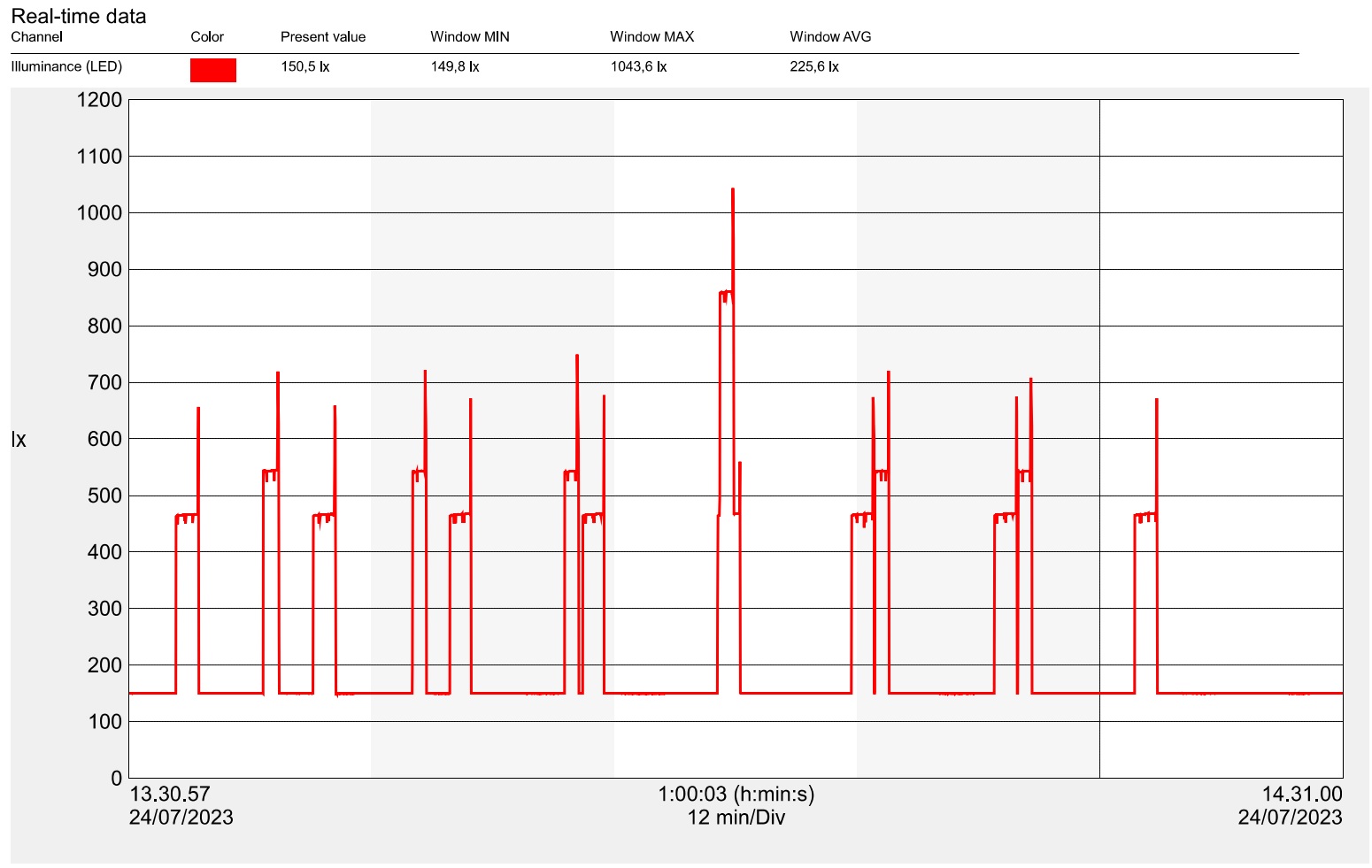}}
\caption{ The lux level variation experienced by the SSN during \textit{T\textsubscript{Int}} for an \textit{N} = 6 configuration under 150 lx illumination, as recorded by the Chauvin Arnoux 111C lux logger software interface. }
\label{LUX_LOG}
\end{figure}

Due to the ability to measure VL using photo-metric lux values, the Chauvin Arnoux 111C lux logger was substituted at the SSN location, maintaining the same orientation as the PV face earlier. This allowed variations in lux around the SSN to be observed during ETX enabled \textit{{T}\textsubscript{Int} }period. A snapshot from lux logger software which describes the variation of lux values during  \textit{{T}\textsubscript{Int}} period is given in Figure \ref{LUX_LOG}.

According to Figure \ref{LUX_LOG}, it is evident that the lux level at the SSN's location varied between 150 lx and 1043.8 lx, maintaining an average value of 225 lx throughout the \textit{{T}\textsubscript{Int}} duration. Consequently, it is apparent that the ETX mode increased the average illumination around the SSN by 50\%. Furthermore, it can be observed that the PSN receiving approximately 1000 lx exhibits an \textit{N} value of exactly 6, while the PSN receiving illumination greater than 1000 lx displays an \textit{N} value greater than 6. Based on the aforementioned results, it can be concluded that ETX has the potential to significantly extend the operational lifespan of the SSN by efficiently sharing surplus energy generated from the PSN nodes. The enabled ETX network demonstrated the ability to operate in a self-sustaining manner continuously for a span of 60 hours. This accomplishment serves as a compelling proof-of-concept for the energy routing and sharing concept of the DE-IoT, as the entire energy rerouting essentially relies on the powered PSN-SSN link pairs, as effectively demonstrated in the results.

\section{Discussion}

This article introduces the novel concept of DE-LIoT, a VLC-based system capable of data and energy networking for densely packed distributed sensor nodes based on energy harvesting (EH). It also discusses the use of existing illumination infrastructure as a centralised controller for distributed LIoT within indoor premises to enable sustainable, energy-autonomous sensor networks. The fundamental mechanism of this network leverages the surplus energy generation capacity of sensor nodes that have more energy resources than they need for their own operations. This is a consequence of the non-uniform distribution of energy/illumination levels in a practical indoor channel. This surplus energy is harnessed to enhance data communication among other nodes through OWC and energy networking via OWPT. The implementation of the prototype OAP and the three nodes for this work followed a less complex, pragmatic approach, utilising commercially available, general-purpose hardware components with possible components made with PE technology. The DE-LIoT proof-of-concept system prototype satisfies engineering requirements for a temperature monitoring application requiring operation under 1000 lux illumination, a 1-hour sensing cycle, and the ability to operate for 8 hours under more than 80\% illumination degradation with limited resources. By incorporating localisation features and omnidirectional transmission and receiving capabilities, this prototype can be further enhanced for future work. Given the commercial unavailability of sustainable electronics, such as PE-based components, the prototype initially employed conventional electronic parts. Nonetheless, with the expected advances in PE-based components technology to mirror conventional electronics, a similar design and performance approach can be adopted for future PE-utilising nodes. Once large-scale, low-cost manufacturing technologies, such as PE and approaches that minimise and reduce all subsystems used for DE-LIoT designs, such as system-on-chip fabrication, are fully developed, these technologies could be used to deploy low-cost, commercially viable DE-LIoT systems in the future.  The prototype implementation successfully utilised consumer device IR communication protocols, which could be a potential approach for large-scale, standardised implementations in the future, with sufficient flexibility for a scaled-up network.  The algorithms employed by the nodes enabled them to effectively communicate all essential parameters to the OAP. Based on this information, the OAP was able to provide feedback and perform energy routing for resources lacking in the SSN node. It is worth noting that, since the static VLC channel conditions remained unchanged and the OAP managed to maintain a LOS throughout the experiments, no inter-node communication was considered in this set-up. While this proof of concept work does not showcase an OAP performing complex optimisation algorithms for large-scale routing, it serves as the foundational building block that forms such systems. 
Considering the scalability of this network, nodes under an OAP can be seen as an individual IoT cluster operating in small indoor area. These clusters can further collaborate and form larger clusters through the establishment of OAP clusters in neighbouring areas. This hierarchical approach can effectively address the scalability challenges of the proposed network. In scenarios where applications demand densely packed smart sensing nodes with limited light ingress, the DE-LIoT system may perform sub-optimally. However, to mitigate this issue, DE-LIoT can be integrated with existing IoT standards such as BLE, forming a hybrid network that leverages both  LOS and NLOS communications. This expands DE-LIoT's reach beyond WPANs, improving scalability and network size.The presented concept extends beyond VLC applications, offering new avenues in energy-efficient networking. This proposed technique supports sustainable wireless networks, aligning with future 6G goals.

\section{Conclusions}

This paper proposes DE-LIoT, a novel concept for a data and energy VLC-based centralised controller and distributed nodes network for EH-based autonomous, densely deployed sensor WPANs. Leveraging OWPT, nodes with higher energy harvesting capabilities share surplus energy, optimising resource utilisation and enhancing network energy efficiency. The paper discusses essential concepts, including data and energy relays and node prioritisation to improve communication capabilities while minimising operational disruptions due to energy shortages. A comprehensive review of relevant DE-LIoT technologies is provided, along with the discussion of proposed network-compatible OAPs and nodes. The performance is evaluated using prototype nodes, demonstrating the concept's potential for unlimited lifetime operation. The findings underscore the impact of the ETX function, the significance of the energy share parameter, and the influence of lux levels on energy acquisition. The feasibility of the evaluated prototype DE-LIoT network transitioning into a commercially viable network has been discussed. The future application of DE-LIoT aims to revolutionise massive sensing and analytics, establishing sustainable, autonomous networks with extended lifespans, enhanced functionality, and increased adaptability.
\section*{Acknowledgment}
This research was supported by the Research Council of Finland (formerly the Academy of Finland) 6G Flagship Programme (Grant Number: 346208), the INDIFICORE , and the SUPERIOT project. The SUPERIOT project has received funding from the Smart Networks and Services Joint Undertaking (SNS JU) under the European Union's Horizon Europe research and innovation programme under Grant Agreement No 101096021, including top-up funding by UK Research and Innovation (UKRI) under the UK government’s Horizon Europe funding guarantee. Views and opinions expressed are however those of the authors only and do not necessarily reflect those of the European Union, SNS JU or UKRI. The European Union, SNS JU or UKRI cannot be held responsible for them.

\bibliographystyle{unsrtnat}







\end{document}